\documentclass[sigconf]{acmart}

\usepackage{tikz}
\usetikzlibrary{fit}
\usetikzlibrary{arrows.meta,positioning,fit,calc}

\tikzset{
  commit/.style={circle,draw,inner sep=1.4pt,minimum size=4mm},
  mainline/.style={-Latex,thick},
  feature/.style={-Latex,thick,dashed},
  label/.style={font=\small},
  buggy/.style={commit,fill=orange!20,draw=orange!60!black},
  merge/.style={commit,fill=blue!15,draw=blue!50!black},
}

\usepackage{siunitx}  
\usepackage{listings}
\definecolor{dkgreen}{rgb}{0,0.6,0}
\definecolor{gray}{rgb}{0.5,0.5,0.5}
\definecolor{mauve}{rgb}{0.58,0,0.82}

\lstdefinestyle{mystyle}{
    language=C++,       
    commentstyle=\color{gray},
    numbers=left,       
    numberstyle=\footnotesize,
    numbersep=5pt,
    frame=single,       
    tabsize=4           
}

\newcommand{\resultbox}[1]{%
    \par\addvspace{6pt plus 2pt minus 1pt}
    \noindent\begin{minipage}{\columnwidth}
        \centering
        \begin{tikzpicture}
            \node [fill=black!10, draw=none, rounded corners=6pt, 
                   inner sep=6pt, outer sep=0pt,
                   text width=0.95\linewidth] (box) {#1};
        \end{tikzpicture}
    \end{minipage}
    \par\addvspace{6pt plus 2pt minus 1pt}
}

\newcommand{\framework}{\textsc{PhantomRun}}

\ccsdesc[500]{Software and its engineering~Software maintenance tools}
\ccsdesc[300]{Software and its engineering~Empirical software validation}
\ccsdesc[100]{Software and its engineering~Software testing and debugging}

\copyrightyear{2026}
\acmYear{2026}
\setcopyright{cc}
\setcctype{by}
\acmConference[MSR '26]{23rd International Conference on Mining Software Repositories}{April 13--14, 2026}{Rio de Janeiro, Brazil}
\acmBooktitle{23rd International Conference on Mining Software Repositories (MSR '26), April 13--14, 2026, Rio de Janeiro, Brazil}
\acmPrice{}
\acmDOI{10.1145/3793302.3793346}
\acmISBN{979-8-4007-2474-9/2026/04}

\begin{document}

\title[\framework: Auto Repair of Compilation Errors in Embedded OSS]{%
  \texorpdfstring{\framework: Auto Repair of Compilation Errors \linebreak in Embedded Open Source Software}{\framework: Auto Repair of Compilation Errors in Embedded Open Source Software}
}

\makeatletter

\author{Han Fu}
\affiliation{%
  \institution{Ericsson AB}
  \city{Stockholm}
  \country{Sweden}
}
\affiliation{%
  \institution{KTH Royal Institute of Technology}
  \city{Stockholm}
  \country{Sweden}
}
\email{hfu@kth.se}

\author{Andreas Ermedahl}
\affiliation{%
  \institution{Ericsson AB}
  \city{Stockholm}
  \country{Sweden}
}
\affiliation{%
  \institution{KTH Royal Institute of Technology}
  \city{Stockholm}
  \country{Sweden}
}
\email{andreas.ermedahl@ericsson.com}

\author{Sigrid Eldh}
\affiliation{%
  \institution{Ericsson AB}
  \city{Stockholm}
  \country{Sweden}
}
\affiliation{%
  \institution{Carleton University}
  \city{Ottawa}
  \country{Canada}
}
\affiliation{%
  \institution{Mälardalen University}
  \city{Västerås}
  \country{Sweden}
}
\email{sigrid.eldh@ericsson.com}

\author{Kristian Wiklund}
\affiliation{%
  \institution{Ericsson AB}
  \city{Stockholm}
  \country{Sweden}
}
\email{kristian.wiklund@ericsson.com}

\author{Philipp Haller}
\affiliation{%
  \institution{KTH Royal Institute of Technology}
  \city{Stockholm}
  \country{Sweden}
}
\email{phaller@kth.se}

\author{Cyrille Artho}
\affiliation{%
  \institution{KTH Royal Institute of Technology}
  \city{Stockholm}
  \country{Sweden}
}
\email{artho@kth.se}

\renewcommand{\shortauthors}{Fu et al.}

\begin{abstract}
Continuous integration (CI) pipelines for embedded software sometimes fail during compilation, consuming significant developer time for debugging. We study four major open-source embedded system projects, spanning over~\num{4000} build failures from the project's CI runs. We find that hardware dependencies account for the majority of compilation failures, followed by syntax errors and build-script issues. Most repairs need relatively small changes, making automated repair potentially suitable as long as the diverse setups and lack of test data can be handled.

In this paper, we present \framework{}, an automated framework that leverages large language models (LLMs) to generate and validate fixes for CI compilation failures. The framework addresses the challenge of diverse build infrastructures and tool chains across embedded system projects by providing an adaptation layer for GitHub Actions and GitLab CI and four different build systems. \framework{} utilizes build logs, source code, historical fixes, and compiler error messages to synthesize fixes using LLMs. Our evaluations show that \framework{} successfully repairs up to 45\,\% of CI compilation failures across the targeted projects, demonstrating the viability of LLM-based repairs for embedded-system CI pipelines.

\end{abstract}

\keywords{continuous integration, software build, large language model, compilation error, program repair} 

\maketitle
\section{Introduction}\label{introduction}
Hardware–software co-development in industrial embedded systems often leads to build failures during continuous integration (CI), particularly in the compilation phase. These failures are common in large organizations and are frequently caused by dependency issues~\cite{fu2022prevalence}. Compilation errors are among the most costly, as they disrupt the entire CI workflow, delay integration, and block testing that depends on successful builds. Conventional repair methods struggle to address such errors in embedded systems, and current software ecosystems lack the infrastructure to support effective automated repair~\cite{DBLP:conf/sigsoft/0001RJGA19}.

In open-source software (OSS) embedded projects, CI pipelines similarly comprise full workflows that include compilation, analysis, and testing. However, this paper focuses exclusively on compilation failures, which surface in diverse forms—ranging from toolchain mismatches to configuration issues—and place a heavy burden on contributors who must diagnose and resolve build breakages. Understanding the distribution of such errors is therefore an essential step toward building automated support that reduces developer effort and accelerates feedback.

To address this challenge, we propose and evaluate \framework{}, an automated repair framework that automatically resolves compilation failures with LLMs without human intervention. We analyze over~\num{10000} pull and merge requests from four major OSS embedded projects (OpenIPC, STM32, RTEMS, and Zephyr)~\footnote{\href{https://github.com/OpenIPC/firmware}{OpenIPC}, \href{https://github.com/platformio/platform-ststm32}{STM32}, \href{https://gitlab.rtems.org/rtems/rtos/rtems}{RTEMS}, \href{https://github.com/zephyrproject-rtos/zephyr}{Zephyr}}, we successfully reproduced \num{4248} compilation errors. 

Our results reveal that hardware-related dependency issues dominate build failures, reflecting an inherent imbalance in embedded CI pipelines. Using \framework{}, we assess the repair capabilities of four OSS LLMs, demonstrating that while code-oriented pretraining improves repair effectiveness, contextual alignment with project-specific examples is equally critical. Most repairs involve small, localized edits, indicating strong potential for practical, low-overhead integration of LLM-based repair into continuous development workflows.

Consequently, we investigate the following research questions:

\textbf{RQ1:}~\textit{What adaptations are needed to make LLM-based repairs applicable to the CI machineries, build infrastructures, and tool chains used by OSS embedded systems?}\label{RQ1}

\textbf{RQ2:}~\textit{How are compilation errors distributed across OSS embedded systems?}\label{RQ2}

\textbf{RQ3:}~\textit{What repair success rates can be achieved with LLM-based repairs for fixing compilation errors in OSS embedded systems?}\label{RQ3}

Our main contributions are:
\begin{enumerate}
  \item \textbf{Framework adaptability:} \framework{} operates seamlessly across diverse CI environments, including four types of build systems in GitHub Actions and GitLab CI/CD.
  \item \textbf{Empirical analysis:} We examine compilation errors in four OSS embedded projects, showing that hardware dependency issues dominate build failures.
  \item \textbf{Repair factors:} We provide insights into how LLM selection and the use of contextual human-fix examples influence automated repair success, achieving a maximum repair success rate of 45\,\%.
\end{enumerate}

This study provides empirical evidence that automated LLM-based repair can effectively resolve real-world compilation failures in embedded CI environments, advancing the automation frontier toward more resilient and self-healing build pipelines.

The remainder of the paper is organized as follows. Section~\ref{sec:background_related_work} introduces the context and motivation of our research and reviews related work on open-source software (OSS) CI reconstruction, hardware-in-the-loop systems with dependency errors, and LLM-based code repair. Section~\ref{sec:study_setup} outlines the study design and research methodology, while Section~\ref{sec:experimental_Adaptation} presents our experimental adaptation addressing RQ1. Section~\ref{sec:compilation_errors_distri} analyzes the distribution of compilation errors across four OSS projects (RQ2), and Section~\ref{sec:LLM-based_repairs} evaluates the repair performance and CI pass rates of \framework{} (RQ3). Section~\ref{sec:threats2validity} discusses threats to validity, Section~\ref{sec:conclusion} summarizes the main findings, and Section~\ref{sec:futurework} concludes future work.

\section{Background and Related work}\label{sec:background_related_work}
Continuous integration (CI) in embedded software development frequently encounters build failures during the compilation phase, where hardware–software co-evolution introduces complex interdependencies. Such failures are especially prevalent in large-scale industrial settings and are often driven by mismatched dependencies, evolving toolchains, or configuration inconsistencies~\cite{fu2022prevalence}. Compilation errors are particularly disruptive, as they halt the CI pipeline, delay integration, and prevent subsequent testing stages that rely on successful builds. Our research builds upon previous work on automated repair of CI failures in industrial environments~\cite{FuEWEHA24}, extending it to the open-source embedded domain. This study aims to bridge the gap between industrial and OSS settings by developing a unified framework that reproduces, analyzes, and automatically repairs compilation failures across diverse CI and build systems.

\subsection{Open-source software (OSS) CI reconstruction}
GitHub Actions (GHA) and GitLab CI are widely used platforms for both industrial and open-source software development, enabling automation of CI workflows.\footnote{\url{https://github.com/features/actions}, \url{https://docs.gitlab.com/ci/}} A Git-based software project may include a CI system that runs automatically on code updates. Each run consists of steps such as building the code for multiple targets (e.g., hardware boards), and a build may fail due to one or more compilation errors with different causes. Both define workflows through YAML configuration files that specify stages and jobs for building, testing, and deploying software. 

In GHA, files like \texttt{build.yml} reside in the \texttt{.github/workflows/} directory and are triggered by repository events such as pushes or pull requests. In contrast, GitLab CI uses a single \texttt{.gitlab-ci.yml} file located at the project root, executed by GitLab runners within customizable environments. While GHA offers seamless integration within the GitHub environment,~\cite{github_actions_about_2024} GitLab CI provides greater flexibility for complex, multi-stage pipelines across self-hosted or cloud infrastructures.~\cite{gitlab_ci_docs_2024}

Recent research has increasingly focused on reconstructing CI builds to enable replicable, data-driven software engineering studies; however, build log analysis is not yet widely used. Several tools have been proposed to enhance reconstructibility in GHA and GitLab CI. ActionRemaker reconstructs and replays workflows to reconstruct specific build conditions~\cite{DBLP:conf/wcre/GolzadehDM22,DBLP:conf/icse/ZhuGFR23}, EGAD~\cite{DBLP:conf/msr/ValenzuelaToledoBKN23} analyzes historical GHA data to identify recurrent workflow issues, and GitBug-Actions~\cite{DBLP:conf/icse/SaavedraSM24} replays bug-fix commits to study CI reliability. Although recent studies have examined GitLab CI pipelines for build failures and efficiency, it still lacks a unified framework capable of processing both GHA and GitLab CI data while supporting automated repair during build execution. In contrast, \framework{} enables systematic CI reconstruction and automated build-stage repair for reproducible failure analysis and resolution.

\subsection{Hardware-in-the-loop with dependency errors}
Although software bugs have been extensively studied, dependency-related failures remain comparatively underexplored. Kerzazi et al.\cite{kerzazi2014automated} and Fu et al.\cite{fu2022prevalence} show that dependency issues account for a substantial share of CI build failures across both OSS and industrial settings. Similar challenges arise in embedded systems, where hardware dependencies and version incompatibilities hinder CI adoption. Olsson et al.\cite{olsson2012climbing} and Lwakatare et al.~\cite{lwakatare2016towards} found that hardware-oriented organizations struggle to apply CI practices effectively, as embedded systems are often user-owned and tightly coupled with proprietary hardware, making automated validation and continuous testing difficult. Prior research~\cite{fischer2020forgotten,khazem2018making,zakaria2022mapping,ratti2018conceptual,DBLP:conf/sigsoft/0001RJGA19} highlights their implications for portability, reconstructibility, and configuration management, yet accurately localizing such errors remains a challenge. Fu et al.~\cite{fu2025autorepairtestcasesllms} analyzed a specific industrial Ericsson system, showing that LLM-equipped CI pipelines can automatically resolve up to 63\,\% of compilation errors, though validation on open-source projects is needed for broader generalization.

\subsection{Log parsing}
Log analysis plays a central role in this area, yet manual log structuring is labor-intensive and difficult to scale~\cite{rodrigues2021clp}. Automated log analysis methods such as LogRAM, DeepLog, and LogTools~\cite{dai2020logram,du2017deeplog,zhu2019tools} have advanced large-scale log parsing but remain underexplored in OSS CI pipelines, where logs are highly diverse and unstructured. Previous work has used log data for execution tracing and root-cause detection~\cite{fu2014digging,he2018identifying,zhang2021onion}, yet systematic studies targeting CI build logs are scarce. Addressing this gap, our study combines log parsing and commit tracing to localize and understand compilation errors across large-scale OSS CI systems, paving the way for automated build failure repair.

\subsection{Automated Program Repair with LLMs}
LLMs have recently become key enablers of Automated Program Repair (APR) due to their strong capacity for understanding and generating source code~\cite{DBLP:journals/corr/abs-2303-18223}. Pre-trained on massive code corpora, transformer-based LLMs capture complex syntactic and semantic relationships, allowing context-aware fixes that surpass traditional heuristic- or template-based approaches~\cite{wei2022chain,PanLWCWW24}. APR has evolved from early genetic-programming systems such as GenProg~\cite{DBLP:journals/tse/GouesNFW12,DBLP:conf/icse/WeimerNGF09} through heuristic-, constraint-, and pattern-based methods~\cite{DBLP:journals/tse/AfzalMSBG21,DBLP:conf/icse/MechtaevYR16} to deep learning (DL)- and LLM-driven repair~\cite{DBLP:journals/tse/ChenKM23,DBLP:conf/icse/FanGMRT23,DBLP:journals/pacmse/Hossain0Z0CLNT24}. However, most existing approaches focus on runtime or test-suite–driven bugs, offering limited applicability to compilation errors in CI pipelines, where repairs must generalize across similar build failures. In this context, LLMs uniquely combine code reasoning and generative capabilities to reconstruct and automatically repair build errors during CI execution. Recent decoder-only models—including Codex~\cite{DBLP:journals/corr/abs-2107-03374}, CodeGen~\cite{DBLP:journals/corr/abs-2305-02309}, CodeLlama~\cite{DBLP:journals/corr/abs-2308-12950}, and DeepSeek-Coder~\cite{DBLP:journals/corr/abs-2401-14196}—demonstrate state-of-the-art performance in code generation, positioning them as core technologies for achieving fully autonomous CI build reconstruction and correction.

\section{Study setup}\label{sec:study_setup}
\begin{figure}[htbp]
    \centering  
    \includegraphics[width=\columnwidth]{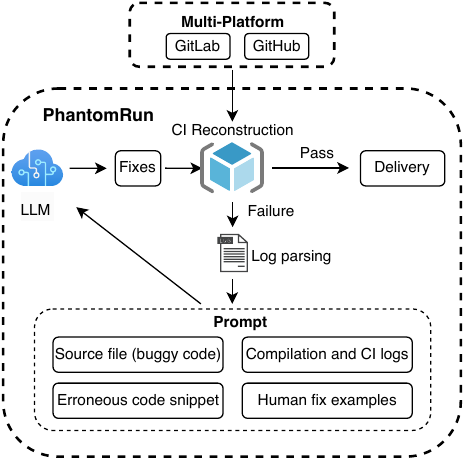}
    \Description{A diagram illustrating the whole process of PhantomRun}
    \caption{PhantomRun process}
    \label{fig:PhantomRun_Process}
\end{figure}

We introduce \framework{}, a fully automated, generative AI framework for repairing compilation errors within CI pipelines, as shown in Fig.~\ref{fig:PhantomRun_Process}. Operating as a non-intrusive, hardware-in-the-loop system, \framework{} analyzes and repairs build failures during each CI cycle without disrupting existing workflows. The framework is designed to operate seamlessly across open-source embedded projects written in C and C++.

Fig.~\ref{fig:PhantomRun_Process} shows the process of CI reconstruction and generating fixes in \framework{}. The framework operates by first reconstructing the CI environment of a failed GitHub pull request (PR) or GitLab merge request (MR), then identifying and extracting the compilation error from the CI logs. Subsequently, \framework{} generates a prompt for an LLM to produce a candidate fix, which is validated by reintroducing it into the CI pipeline. This process iterates until a successful build is achieved or a maximum number of attempts is reached. We set the maximum number of attempts to five, which is sufficient to cover most compilation errors and prevents potential infinite repair loops~\cite{fu2025autorepairtestcasesllms}.

\subsection{Project selection and data collection}\label{subsec:project_data_collection}
Table~\ref{tab:CI_Tooling} summarizes the four projects analyzed in this study. For the open-source component, repositories were collected from GitHub using authenticated API queries filtered by three criteria: (1) tagged with \texttt{embedded}, (2) primarily written in C or C++, and (3) having more than 20 stars. These filters produced about 60 candidates, from which Zephyr, OpenIPC, and STM32 were selected for detailed analysis. The \texttt{embedded} tag ensured relevance to embedded systems, language restrictions reflected the dominance of C/C++ in embedded development, and the star threshold served as a proxy for project maturity. Additional metadata, such as activity and popularity, guided the final selection, yielding a curated, diverse dataset suitable for automated analysis and comparison. In addition, as with GitHub, we conducted automated exploration of public projects on GitLab instances, including the RTEMS GitLab server. We collected data through automated queries to the GitHub and GitLab APIs, focusing on pull and merge requests (PRs and MRs) that failed during the build stage. Data collection begins with the first PR or MR for each project through October 3, 2025; therefore, later PRs or MRs are not included. 

\begin{table*}[t]
\centering
\caption{Comparison of CI Tooling, Infrastructure, and Static Analysis across selected projects.}
\label{tab:CI_Tooling}
\renewcommand{\arraystretch}{1.4}
\setlength{\tabcolsep}{4pt}
{\small
\begin{tabular}{@{}>{\raggedright\arraybackslash}p{2.6cm}p{3.2cm}<{\raggedright}p{3.2cm}<{\raggedright}p{3.2cm}<{\raggedright}p{3.2cm}<{\raggedright}@{}}
\toprule
\textbf{Aspect} & \textbf{OpenIPC Firmware} & \textbf{PlatformIO STM32} & \textbf{RTEMS} & \textbf{Zephyr} \tabularnewline
\midrule
\textbf{CI Platforms} 
& GitHub Actions; GitHub pull requests
& GitHub Actions; GitHub pull requests
& GitLab CI/CD; GitLab merge requests
& GitHub Actions; GitHub pull requests \tabularnewline

\textbf{OS / Language} 
& Linux-based / C  
& Multi-OS / Python, C++ 
& RTOS / C  
& RTOS / C, C++    \tabularnewline

\textbf{Community} 
& Enthusiast and hacker community
& Broad embedded community (PlatformIO users)
& Expert-focused
& 1600+ contributors, large community    \tabularnewline 

\textbf{Build failures in six months} 
& 31
& 444
& 1784
& 1989    \tabularnewline 

\textbf{Build Tools} 
& \texttt{make} + Buildroot
& Python (SCons)
& \texttt{make} + autotools
& \texttt{west} + CMake + Ninja \tabularnewline

\textbf{Infrastructure} 
& Relies on SoC vendor SDKs and custom toolchains  
& Python; JSON-based board configs; PlatformIO build scripts
& GitLab instance; custom build system  
& CMake + Kconfig; modular kernel; managed with \texttt{west}  \tabularnewline

\textbf{Log Structure} 
& Aggregated (Make $\rightarrow$ Buildroot $\rightarrow$ cross-GCC/LD)
& Layered (PlatformIO/SCons $\rightarrow$ toolchain (GCC/Clang) $\rightarrow$ LD)
& Autotools (configure $\rightarrow$ make $\rightarrow$ GCC/LD)
& Structured (CMake $\rightarrow$ Ninja $\rightarrow$ GCC/LD) \tabularnewline

\textbf{Typical Failure Sources} 
& Toolchain setup, packaging, kernel config
& Python invocation, SCons tasks, compiler/linker
& Autoconf macros, Makefile rules, linker scripts
& CMake/Kconfig/DTS, compiler, linker     \tabularnewline 

\textbf{Error Pattern Coverage} 
& Make, Buildroot, cross-compiler, packaging
& Compiler, linker, SCons, Python exceptions
& Compiler, linker, configure, make
& Compiler, linker, CMake, Ninja, Kconfig, west    \tabularnewline 

\textbf{Example Error} 
& \texttt{make[2]: *** [package] Error 2}
& \texttt{[env:esp32] src/main.cpp:12:10: fatal error: Arduino.h not found}
& \texttt{arm-rtems6-gcc: error: undefined reference to 'rtems\_task\_create'}
& \texttt{src/main.c:42:17: error: 'foo' undeclared}    \tabularnewline 

\bottomrule
\end{tabular}
}
\renewcommand{\arraystretch}{1.0}
\end{table*}

\subsection{LLM selection and prompt design}\label{subsec:LLMSelection_Prompt_Design}
\begin{table}[t]
  \centering
  \caption{Large Language Models}
  \label{tab:LLMs}
  \renewcommand{\arraystretch}{1.2}
  \setlength{\tabcolsep}{6pt}
{\small
  \begin{tabular}{@{}lcc@{}}
    \toprule
    \textbf{Model} & \textbf{Trained on Source Code} & \textbf{Parameter Size} \tabularnewline
    \midrule
    CodeT5+   & Yes & 7B \tabularnewline
    CodeLlama & Yes & 7B \tabularnewline
    Falcon    & No  & 7B \tabularnewline
    Bloom     & No  & 7B \tabularnewline
    \bottomrule
  \end{tabular}
}
\end{table}
We integrated four large language models (LLMs) into the repair process and evaluated their performance using two metrics: \emph{accuracy}, the proportion of verified successful fixes, and \emph{efficiency}, the reduction in repair effort based on fix size and quality. The selected models (Table~\ref{tab:LLMs}), including CodeT5+, CodeLlama, Falcon, and Bloom, were chosen for accessibility, code-oriented training, dataset diversity, and suitability for local deployment. All models run are limited to 7B parameters to ensure efficient inference within CI workflows.

\begin{figure}[htbp]
    \centering  
    \includegraphics[width=\columnwidth]{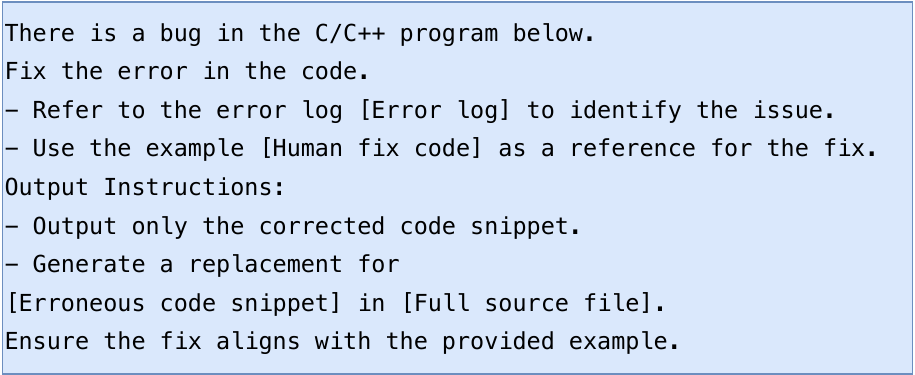}
    \Description{A diagram showing the prompt to generate fixes.}
    \caption{Prompt template used to generate fixes}
    \label{fig:Prompt}
\end{figure}

Prompt engineering has emerged as a crucial technique for enhancing the capabilities of pre-trained LLMs~\cite{sahoo2025systematicsurveypromptengineering}. In our study, prompts that include the source file, compilation and CI logs, the erroneous code snippet, and examples of human fixes enable LLMs to generate more accurate repairs for compilation errors. Fig.~\ref{fig:Prompt} illustrates such a prompt template, which instructs the model to produce a corrected code segment to replace the \texttt{[Erroneous code snippet]} within the \texttt{[Full source file]}.

\subsection{Terminology}\label{subsubsec:terminology}
In this study, we define the key terms used in the prompt template as follows:
\begin{itemize}
    \item \textit{Buggy baseline.} The exact commit on main right before the PR or MR’s changes first become part of main. 
    \item \textit{Erroneous code snippet.} The error code extracted based on the CI log.   
    \item \textit{Human fix example} refers to a specific segment of a commit that resolves a compilation error. 
    \item \textit{Lines of code (LOC)} refers to the number of lines modified in a successful repair.
\end{itemize}

A \emph{segment} refers to one or more lines of code that may not form a complete syntactic unit. The \emph{erroneous code snippet} denotes the portion that triggers a compilation error, as identified in CI logs, and serves as the focal point for generating fixes. To guide the repair process, we derive \emph{human fix examples}—code segments from commits that resolved similar compilation errors—by locating the commit ID, faulty file, and line positions from failed builds, then comparing them with the subsequent successful build to isolate the corrective change. 

\emph{Lines of code (LOC)} measure the number of lines modified in a successful repair. A size of 0 denotes changes unrelated to source content, such as permission updates, binary modifications, or complete file deletions. A size of 1 represents a single-line addition or deletion, while a size of 2 typically corresponds to a one-line modification (one addition and one deletion) or two added or removed lines. Treating full file removals as 0-LOC changes prevents inflating metrics due to large-scale deletions sometimes produced by LLM-generated fixes that merely bypass CI verification rather than correct the underlying issue.

\subsubsection{Identifying failed commits}\label{subsubsec:identifying_failed_commits}
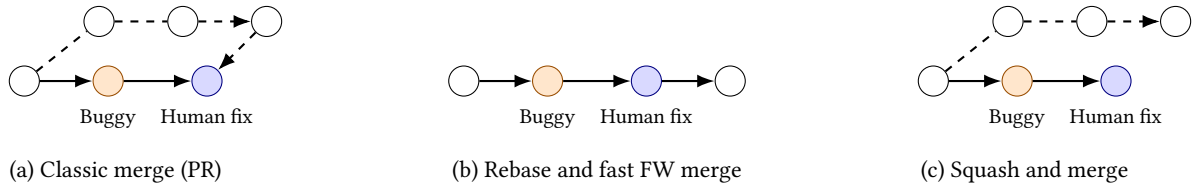
\begin{figure*}[t]
\centering
\begin{minipage}[t]{0.32\textwidth}
\centering
\begin{tikzpicture}[node distance=7mm]
\node[commit,label={[label]below:A1}] (A1) {};
\node[commit,right=of A1,buggy,label={[label]below:A2}] (A2) {};
\draw[mainline] (A1) -- (A2);

\node[commit,above right=5mm and 7mm of A1,label={[label]above:B1}] (B1) {};
\node[commit,right=of B1,label={[label]above:B2}] (B2) {};
\node[commit,right=of B2,label={[label]above:B3}] (B3) {};
\draw[feature] (A1) -- (B1) -- (B2) -- (B3);

\node[merge,right=9mm of A2,label={[label]below:M}] (M) {};
\draw[mainline] (A2) -- (M);
\draw[feature] (B3) -- (M);

\node[below=.5mm of A2,label] {Buggy};
\node[below=.5mm of M,label] {Human fix};
\node at ($(A1)!0.5!(M)+(0,-1.2)$) {{(a) Classic merge (PR)}};
\end{tikzpicture}
\end{minipage}
\hfill
\begin{minipage}[t]{0.32\textwidth}
\centering
\begin{tikzpicture}[node distance=7mm]
\node[commit,label={[label]below:A1}] (A1) {};
\node[commit,right=of A1,buggy,label={[label]below:A2}] (A2) {};
\draw[mainline] (A1) -- (A2);

\node[commit,merge,right=9mm of A2,label={[label]below:B1'}] (B1) {};
\node[commit,right=of B1,label={[label]below:B2'}] (B2) {};
\draw[mainline] (A2) -- (B1);
\draw[mainline] (B1) -- (B2);

\node[below=.5mm of A2,label] {Buggy};
\node[below=.5mm of B1,label] {Human fix};
\node at ($(A1)!0.5!(B2)+(0,-1.2)$) {{(b) Rebase and fast FW merge}};
\end{tikzpicture}
\end{minipage}
\hfill
\begin{minipage}[t]{0.32\textwidth}
\centering
\begin{tikzpicture}[node distance=7mm]
\node[commit,label={[label]below:A1}] (A1) {};
\node[commit,right=of A1,buggy,label={[label]below:A2}] (A2) {};
\draw[mainline] (A1) -- (A2);

\node[commit,above right=5mm and 7mm of A1,label={[label]above:B1}] (B1) {};
\node[commit,right=of B1,label={[label]above:B2}] (B2) {};
\node[commit,right=of B2,label={[label]above:B3}] (B3) {};
\draw[feature] (A1) -- (B1) -- (B2) -- (B3);

\node[merge,right=9mm of A2,label={[label]below:S}] (S) {};
\draw[mainline] (A2) -- (S);

\node[below=.5mm of A2,label] {Buggy};
\node[below=.5mm of S,label] {Human fix};
\node at ($(A1)!0.5!(S)+(0,-1.2)$) {{(c) Squash and merge}};
\end{tikzpicture}
\end{minipage}
\Description{A diagram showing how to find the erroneous code.}
\caption{Identification of buggy baseline and human fix for different Git merge strategies}
\label{fig:merge-strategies-buggy-baseline}
\end{figure*}

Figure~\ref{fig:merge-strategies-buggy-baseline} illustrates how erroneous and fixed code segments are derived under different integration strategies, including merged, rebased, and squashed commits. To collect these examples in OSS projects, we analyze commit histories and CI outcomes to trace build failures and their resolutions. A commit introducing a build failure is labeled as buggy, while the earliest subsequent commit restoring a successful build is identified as the human fix. The buggy baseline depends on the merge strategy: in \textit{classic merges} (see Fig.~\ref{fig:merge-strategies-buggy-baseline} (a)) it is the main branch commit preceding the merge; in \textit{rebase or fast-forward integrations} (see Fig.~\ref{fig:merge-strategies-buggy-baseline} (b)) it is the last main commit before rebasing; and in \textit{squash merges} (see Fig.~\ref{fig:merge-strategies-buggy-baseline} (c)) it is the parent of the squash commit representing the aggregated changes. In all cases, the fixed state corresponds to the commit where the PR or MR is integrated into the main branch, ensuring a consistent definition of the buggy baseline as the last main commit before integration.

\subsection{Statistical Analysis}\label{subsec:statistical_analysis}
To test our hypotheses, we applied two statistical methods—the Chi-Square Test and the Friedman Test—to analyze repair success rates and error distributions across the four projects. These tests were used to assess the significance of differences in repair outcomes and related metrics derived from multiple \framework{} executions. 

\subsubsection{Chi-Square Test}
The Chi-square test is a non-parametric statistical method used to evaluate whether two categorical variables are independent or significantly associated~\cite{mchugh2013chi}. It is widely applied to contingency tables to determine whether observed frequencies differ from expected frequencies under the assumption of independence. Under the null hypothesis $H_0$, the two categorical variables are assumed to be independent. The calculated $\chi^2$ statistic approximately follows a Chi-square distribution with $(r-1)(c-1)$ degrees of freedom. If the computed statistic exceeds the critical value for a chosen significance level (typically $\alpha = 0.05$), $H_0$ is rejected, indicating a statistically significant association between the variables.

\subsubsection{Friedman test}
The Friedman test is a non-parametric alternative to repeated-measures ANOVA~\cite{friedman1937use}, designed to detect differences in treatments across multiple related conditions without assuming normality of the data. It is particularly suitable when the same set of subjects (e.g., CI projects) is evaluated under different conditions (e.g., LLMs). Each observation is first ranked within its subject, and $R_j$ denotes the sum of ranks for condition $j$. A higher $\chi^2_F$ value indicates stronger evidence that at least one condition performs differently from the others. The statistic approximately follows a Chi-square distribution with $k - 1$ degrees of freedom when $n$ and $k$ are reasonably large. 

\subsection{Methodology of RQ1}
To address RQ1, in Section~\ref{sec:experimental_Adaptation}, we adapted \framework{} to reconstruct CI environments across four OSS embedded projects with heterogeneous platforms, build tools, and infrastructures. 

\subsection{Methodology of RQ2}
This research question investigates the prevalence of build failures and types of compilation errors across the four projects, as detailed in Section~\ref{sec:compilation_errors_distri}. We applied the Chi-square test of independence mentioned above to assess whether the distribution of PR or MR failures at the build stage is independent of projects. Formally, the hypotheses are stated as follows: $H_1^0$: Build-stage failure is independent of the project; $H_1^1$: Build-stage failure depends on the project. 

We are also analyzing their build logs and CI pipelines, and we aim to characterize the error distribution (e.g., Environment Setup Failure, Syntax Error, Hardware Dependency Error, Non-hardware Dependency Error, and Compiler Configuration Error). Understanding this distribution provides insight into the typical challenges developers face across different embedded ecosystems and highlights project-specific weaknesses.

\subsection{Methodology of RQ3}
This research question investigates the pass rates of different LLMs within \framework{} when repairing reconstructed CI compilation failures and their effects, as detailed in Section~\ref{sec:LLM-based_repairs}. We first evaluate pass rates using human fix examples in the prompt drawn from projects other than the one under study, then using randomly selected examples across all projects, and finally using examples from the same project being evaluated.

To determine whether the repair rate is influenced by the source of human fix examples, we apply the Friedman test, as the examples are tied to the prompts provided to the LLMs. We compare three related experimental conditions; therefore, we have three ranks (1, 2, 3). Rank 1 = lowest pass rate, Rank 3 = highest pass rate. The hypotheses are defined as: $H_2^0$: The repair rate is independent of the source of human-fix examples; $H_2^1$: The repair rate depends on the source of human-fix examples.

We further examine whether different LLMs significantly affect the CI pass rate using the Chi-square test, comparing their relative repair performance across projects. The hypotheses are: $H_3^0$: The CI pass rate is independent of the LLM; $H_3^1$: The CI pass rate depends on the LLM.

Finally, to assess the influence of compilation error types on repair outcomes, we again apply the Chi-square test. The hypotheses are: $H_4^0$: The repair rate is independent of the build failure (error) type; $H_4^1$: The repair rate depends on the build failure (error) type.

We further assess the quality of repairs generated by \framework{}. Each fix is evaluated by its size, measured in modified LOC, to capture its minimality and precision. Additionally, a qualitative analysis of representative fixes examines whether the changes align with developer intent, preserve functional correctness, and avoid superficial edits that merely bypass CI checks.

\section{Adaptions for CI builds}\label{sec:experimental_Adaptation}
To answer RQ1, this section outlines the necessary adaptations to enable the LLM-based repair applicable to the CI machinery. 

\begin{figure}[htbp]
    \centering  
    \includegraphics[width=\columnwidth]{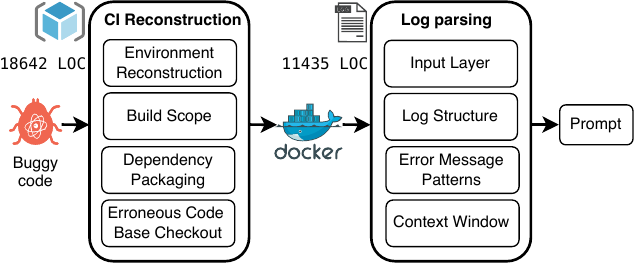}
    \Description{A diagram showing the process of how to reproduce a CI failed run.}
    \caption{CI reconstruction and log parsing}
    \label{fig:CI_Docker}
\end{figure}

The CI reconstruction and log parsing processes are core components of \framework{}. As illustrated in Fig.~\ref{fig:CI_Docker}, the process consists of two components: CI reconstruction and log parsing. After a fix is generated, \framework{} re-executes the CI workflow using the same operating environment, compiler, toolchain, and dependencies as the failed run, thereby guaranteeing that each fix is evaluated under identical conditions to the original failure. The implementation comprises~\num{30077} lines of code, of which~\num{18642} lines implement CI reconstruction, including CI configuration parsing, environment instantiation, build stage extraction, and containerized execution. The remaining~\num{11435} lines support log parsing and analysis, covering compiler diagnostics extraction, error normalization, and rule based classification.

\subsection{CI Reconstruction}\label{subsec:ci_reconstruct}
The \emph{CI reconstruction} component ensures that automatically generated fixes are validated within their original build environments. The process begins by systematically collecting project metadata and verifying the presence of CI files and supporting documentation. Projects lacking a default branch or sufficient metadata are excluded to maintain reproducibility. 

CI reconstruction proceeds through four main steps: (1) recreating the environment of the original PR or MR, (2) selecting the relevant CI scope—specifically, the build stage in our context, (3) identifying dependency packages and external toolchains, and (4) checking out the erroneous code base for faithful reproduction. 

Listing~\ref{lst:ci-build-stage} illustrates the CI reconstruction process, showing how a build stage is extracted from the original CI specification and reproduced in an isolated environment. We reuse the original CI files (e.g., \texttt{.github/workflows/*.yml}, \texttt{.gitlab-ci.yml}) and statically reconstruct build environments by parsing CI specifications and referenced scripts using deterministic pattern matching. From these specifications, we extract the base operating system, container images, environment variables, setup commands, and build steps, and translate them into executable reconstruction artifacts, namely a \texttt{Dockerfile} and a build script. We identify compiler invocations and toolchain provisioning commands by scanning build scripts with regular expression patterns and re execute the same build stage in isolation.

We rely on machine readable metadata in CI specifications to identify base operating or container images, including image names, tags, and, when available, immutable digests, which determine the default compilers, system libraries, and preinstalled tools. To mitigate \emph{environment drift} caused by mutable tags or updated base operating images, we record image identifiers and execution timestamps. Embedded CI pipelines frequently provision cross compilation toolchains and language runtimes via scripted downloads or CI actions; we extract these steps verbatim and record the toolchain family, version, activation method, and all non-secret environment variables, including board selectors, configuration options, compiler flags, and SDK paths.

\begin{lstlisting}[language=bash,basicstyle=\ttfamily\footnotesize,frame=single,caption={Illustration of CI reconstruction},label={lst:ci-build-stage}]
git checkout $CI_COMMIT_SHA
BUILD_CMD=$(parse_ci ci.yml --job build)
# e.g., cmake .. && make | ninja | west build -b $BOARD | pio run
docker run --rm ci-repro $BUILD_CMD
\end{lstlisting}

We restrict CI reconstruction to the \emph{build stage}, defined as the set of CI jobs and steps responsible for compiling the code. We identify build stage jobs by the presence of compiler invocations in CI configurations and logs (e.g., \texttt{cmake}, \texttt{make}, \texttt{ninja}, \texttt{west build}, or \texttt{pio run}). This stage may include preparatory steps required to reach compilation, such as workspace initialization or toolchain setup, but excludes downstream stages that operate on already built artifacts. Within the build stage, we reproduce the build by checking out the commit referenced by the CI job and mirroring repository checkout settings, such as fetch depth, submodules, or workspace managers (e.g., \texttt{west}).

The three GitHub hosted projects reuse a shared, project agnostic CI reconstruction implementation that parses CI configurations, isolates the build stage, and generates deterministic \texttt{Dockerfile}s and build scripts. Project-specific differences are handled through lightweight adapters for build invocation and toolchain setup, while the core logic remains unchanged. The GitLab-hosted RTEMS project uses a separate backend tailored to GitLab CI semantics but follows the same reconstruction methodology, enabling consistent and reproducible CI execution and comparable build log generation across all studied projects.

\subsection{Log parsing}\label{subsec:log_parsing}
Fig.~\ref{fig:CI_Docker} illustrates our log parsing component, which operates across four abstraction layers: the \emph{input layer}, the \emph{log structure layer}, the \emph{error message patterns layer}, and the \emph{context window layer}.

\subsubsection{Input Layer.}
After reconstructing the CI environment, the \emph{input layer} ingests raw logs from \texttt{build\_log.json} files and prepares them for analysis. Logs are decoded using a tolerant strategy (e.g., Python’s \texttt{errors='replace'} or \texttt{errors='ignore'}) to handle mixed or malformed encodings that may result from toolchain outputs, colored terminals, or locale-specific messages. This approach ensures uninterrupted parsing by replacing or skipping invalid bytes while preserving the log structure. The text is then normalized by removing ANSI color codes, progress-bar artifacts, and backspaces, producing a clean and uniform stream for subsequent pattern matching, as illustrated in Fig.~\ref{fig:CI_Docker}.

\subsubsection{Log Structure.}
The \emph{log structure layer} maps each project's build system and execution flow to the compiler or linker diagnostics that reveal the cause of failure. Each build produces distinct log patterns: Zephyr logs begin with the \texttt{west} command orchestrating the CMake--Ninja pipeline; RTEMS emits sequential \texttt{configure} and \texttt{make} traces; OpenIPC nests \texttt{make} and Buildroot invocations; and PlatformIO interleaves Python~(\texttt{SCons}) outputs with compiler messages. By parsing these structures, we isolate the compilation phase—where GCC or Clang report syntactic or semantic errors—while filtering out earlier configuration or packaging steps. The parser identifies compiler entry points from tool-specific prefixes (e.g., \texttt{ninja:}, \texttt{make[2]:}, \texttt{west build:}) and applies corresponding patterns to extract structured diagnostics. The resulting normalized segments yield a unified, compiler-aware representation of build failures, forming the basis for cross-project error analysis and automated repair validation.

\subsubsection{Error Message Patterns.}
The \emph{error message layer} extracts diagnostics from the compilation phase using hierarchical regular expressions that capture common patterns in compiler, linker, and build-tool outputs. Patterns are applied from specific to generic, starting with structured compiler messages referencing files and line numbers, then progressing to unstructured \texttt{Error:} lines—to minimize false positives and prioritize detailed diagnostics. GCC and Clang emit standardized error formats, while linkers (\texttt{ld} or \texttt{ld.lld}) report undefined references or multiple definitions. Build tools such as CMake, Ninja, Make, and SCons produce configuration or task-failure indicators, handled through supplementary patterns. Project-specific adapters extend the core set to accommodate distinct ecosystems: \texttt{west}, \texttt{Kconfig}, and device-tree errors for Zephyr; \texttt{make} and Buildroot for OpenIPC; \texttt{SCons} and Python traces for PlatformIO; and \texttt{configure} or \texttt{make} traces for RTEMS. Each detected error yields a structured record containing the tool type, file, and line information, and a normalized diagnostic message, enabling consistent error extraction across diverse CI environments.

\subsubsection{Compilation Error Distribution.} 
We classify compilation errors by analyzing compiler error messages, build system outputs, and CI job logs. For each failing build, we classify the \emph{first fatal error} that terminates the build to avoid cascading failures. We extract this error by normalizing logs and isolating the compiler or build-system message responsible for stopping the build.

We assign each error to exactly one category using deterministic rules that combine information from the error message and its execution context. We match error messages against predefined regular expression patterns, infer the failing build phase from the surrounding build commands (e.g., configuration, compilation, or linking), and extract referenced files, include paths, or configuration options from both the error message and the command line that triggered it. We apply the rules in a fixed priority order, and the first matching rule determines the category. Based on this procedure, we define five categories:

\begin{itemize}
  \item \textbf{Environment setup failure}: missing build tools or SDKs, failed package installation, or incorrect base operating or container initialization.
  \item \textbf{Syntax error}: language-level compilation errors caused by invalid source code constructs, independent of target hardware.
  \item \textbf{Hardware dependency error}: board, architecture, or System on a Chip (SoC)-specific failures, such as missing device headers, incompatible board support packages, or invalid target configurations.
  \item \textbf{Non-hardware dependency error}: missing or incompatible software dependencies unrelated to hardware, including absent libraries or unresolved third party symbols.
  \item \textbf{Compiler configuration error}: failures due to incorrect compiler flags, incompatible optimization options, or misconfigured toolchain settings.
\end{itemize}

To ensure robustness, we iteratively refined the pattern set through manual inspection of representative samples from each project across different toolchains and build systems. We refined the rules until repeated application produced stable category assignments, with no further rule changes required for newly inspected samples. Because classification relies on deterministic pattern matching and fixed rule priority, reapplying the same rules to the same logs always yields identical results, ensuring consistent and reproducible categorization across the dataset.

\subsubsection{Context Window and Ordering.}
For each matched diagnostic, the parser retains up to two surrounding lines (\texttt{context\_before} $\leq$ 2, \texttt{context\_after} $\leq$ 2) to preserve local context and capture the command or output associated with the error. This information helps reconstruct the build sequence and distinguish compiler faults from higher-level build or configuration errors. Patterns are evaluated in a prioritized order—from specific expressions linked to compilers, linkers, and build tools (e.g., GCC/Clang, \texttt{ld}, CMake/Ninja, autotools, Kconfig, SCons, Buildroot) to generic \texttt{Error:} messages—ensuring that structured diagnostics are captured first and reducing false positives across all four build systems. 

\resultbox{
    \textbf{Finding 1:} Extending the setup to new projects requires only minor adjustments to accommodate project-specific build systems, such as updating configuration files.
}

\section{Compilation Error Distribution}\label{sec:compilation_errors_distri}
To answer RQ2, we analyze the most common types of failures occurring during the build stage. Table~\ref{tab:Data_set_distribution} shows the distribution of pull and merge requests (PRs and MRs) that failed at this stage and the corresponding compilation errors. On average, 10.0\,\%, and 11.25\,\% standard deviation of all CI failures occur at the build stage, with Zephyr showing the lowest failure rate at 8.0\,\%. 

\begin{table}[t]
  \caption{Build failures and compilation errors distribution}
  \centering
  \renewcommand{\arraystretch}{1.2}
  \resizebox{\columnwidth}{!}{%
  \begin{tabular}{l S[table-format=5.0] S[table-format=4.0] S[table-format=2.1, table-space-text-post=\,\%] S[table-format=4.0]}
  \toprule
  \textbf{Project} & {\textbf{Failed PRs/MRs}} & {\textbf{Failed PRs/MRs}} & {\textbf{Build stage}} & {\textbf{Compilation}} \tabularnewline
  & {\textbf{Total}} &  {\textbf{Build stage}} & {\textbf{\%}} & {\textbf{Errors}}\tabularnewline
  \midrule
  OpenIPC          & 114   & 10    & 8.7\,\%   & 31    \tabularnewline
  STM32            & 155   & 40    & 25.8\,\%  & 444   \tabularnewline
  RTEMS            & 761   & 225   & 29.5\,\%  & 1784  \tabularnewline
  Zephyr           & 8984  & 725   & 8.0\,\%   & 1989  \tabularnewline
  \midrule
  \textbf{Total (SD)} & 10014 & 1000  & {10.0 $\pm$ 11.25\,\%}  & 4248 \tabularnewline
  \bottomrule
  \end{tabular}
  }
  \label{tab:Data_set_distribution}
\end{table}

To examine whether build-stage failure rates differ significantly across projects, we conducted a Chi-square test of independence. 
The result ($\chi^2(2) = 402.6128,\, p = 0.0000$) indicates that build-stage failure likelihood varies significantly across projects, reflecting project-specific CI configurations and hardware environments that influence compilation stability.\footnote{Detailed experimental data can be found in the companion artifact repository}

The distribution of the exact types of compilation errors is of particular interest, as shown in Table~\ref{tab:compilationPrevalenceOpenMapped}. Dependency-related errors account for 71.5\,\%, of all failures, with hardware-related issues dominating (61.2\,\%). These typically arise from missing or incompatible BSPs, SDKs, or peripheral libraries across multiple hardware targets. STM32 shows the highest share (77\,\%) due to frequent SDK updates and diverse MCU variants, followed by Zephyr (69\,\%) with its rapid hardware integration. RTEMS experiences fewer such issues owing to its curated support set, while OpenIPC shows the least, reflecting its narrower and more stable hardware base. Overall, the data reveal a clear imbalance, with hardware dependencies being the primary source of build instability.

\begin{table*}[t]
  \caption{Distribution of Compilation Errors}
  \centering
  \renewcommand{\arraystretch}{1.2}
  \setlength{\tabcolsep}{7.6pt}
{\small
  \begin{tabular}{l r@{\hspace{4pt}}r r@{\hspace{4pt}}r r@{\hspace{4pt}}r r@{\hspace{4pt}}r @{\hskip 28pt} r}
  \toprule
  \textbf{Compilation error type} & \multicolumn{8}{c}{\textbf{Occurrences (\%)}} & \textbf{Total (\%)} \tabularnewline
  & \multicolumn{2}{c}{OpenIPC} & \multicolumn{2}{c}{STM32} & \multicolumn{2}{c}{RTEMS} & \multicolumn{2}{c}{Zephyr} & \tabularnewline
  \midrule
  Environment Setup Failure      & 4  & (10\,\%)  & 23  & (5\,\%)   & 286  & (16\,\%)  & 100  & (5\,\%)   & 9.7\,\% \tabularnewline
  Syntax Error                   & 2  & (7\,\%)   & 80  & (18\,\%)  & 321  & (18\,\%)  & 258  & (13\,\%)  & 15.6\,\% \tabularnewline
  Hardware Dependency Error      & 14 & (47\,\%)  & 341 & (77\,\%)  & 874  & (49\,\%)  & 1372 & (69\,\%)  & 61.2\,\% \tabularnewline
  Non-hardware Dependency Error  & 11 & (36\,\%)  & \multicolumn{2}{c}{--} & 249  & (14\,\%)  & 179  & (9\,\%)   & 10.3\,\% \tabularnewline
  Compiler Configuration Error   & \multicolumn{2}{c}{--} & \multicolumn{2}{c}{--} & 54   & (3\,\%)   & 80   & (4\,\%)   & 3.2\,\% \tabularnewline
  \hline
  \textbf{Total}                 & 31 & \phantom{(00\,\%)} & 444 & \phantom{(00\,\%)} & 1784 & \phantom{(00\,\%)} & 1989 & \phantom{(00\,\%)} & 4248 \tabularnewline
  \bottomrule
  \end{tabular}
}
  \label{tab:compilationPrevalenceOpenMapped}
\end{table*}

Syntax errors represent the second most common failure type (15.6\,\%), typically arising from typos, missing semicolons, or incorrect language constructs. Compiler configuration errors are the least frequent (3.2\,\%). Overall, the strong presence of hardware-related dependency issues highlights the unique complexity of embedded CI systems compared to other OSS projects~\cite{DBLP:conf/msr/RauschHL017}. 

\resultbox{
    \textbf{Finding 2:} Most embedded CI build failures stem from hardware dependencies rather than syntax errors, underscoring the need for stronger dependency and configuration management in both industrial and OSS pipelines.
}

\section{LLM-based Repairs}\label{sec:LLM-based_repairs}
To answer RQ3, we analyze the pass rate of different LLMs in \framework{} on the collected CI compilation failures. We use each LLM to interact within \framework{} up to five times to generate and validate fixes. We evaluate three scenarios to assess the effect of human fix examples: (1) randomly selected fix examples across all projects, (2) examples from the same project, and (3) examples from other projects. Finally, we evaluate two erroneous codes and fixes generated by \framework{}. 

\subsection{LLM-based pass rate}\label{sec:LLM-based_passrate}
\begin{table*}[t]
    \caption{Successful compilations (\%) for different combinations of LLMs and example types}
    \centering
    \small
    \begin{tabular}{l *{12}{r}}
    \toprule
    \textbf{LLMs} & \multicolumn{4}{c}{\textbf{Other projects}} & \multicolumn{4}{c}{\textbf{Random human fix}} & \multicolumn{4}{c}{\textbf{Same project}} \tabularnewline
    \cmidrule(lr){2-5} \cmidrule(lr){6-9} \cmidrule(lr){10-13}
     & \multicolumn{1}{c}{OpenIPC} & \multicolumn{1}{c}{STM32} & \multicolumn{1}{c}{RTEMS} & \multicolumn{1}{c}{Zephyr}
     & \multicolumn{1}{c}{OpenIPC} & \multicolumn{1}{c}{STM32} & \multicolumn{1}{c}{RTEMS} & \multicolumn{1}{c}{Zephyr}
     & \multicolumn{1}{c}{OpenIPC} & \multicolumn{1}{c}{STM32} & \multicolumn{1}{c}{RTEMS} & \multicolumn{1}{c}{Zephyr} \tabularnewline
    \midrule
    CodeT5+   & 28 & 31 & 29 & 33 & 30 & 31 & 28 & 33 & 32 & 35 & 33 & 37 \tabularnewline
    CodeLlama & \textbf{38} & \textbf{39} & \textbf{31} & \textbf{41} & \textbf{38} & \textbf{39} & \textbf{35} & \textbf{41} & \textbf{42} & \textbf{43} & \textbf{35} & \textbf{45} \tabularnewline
    Falcon     & 28 & 29 & 25 & 30 & 28 & 29 & 27 & 32 & 31 & 33 & 29 & 33 \tabularnewline
    Bloom      & 27 & 28 & 24 & 28 & 27 & 26 & 25 & 29 & 30 & 31 & 28 & 32 \tabularnewline
    \bottomrule
    \end{tabular}
    \label{tab:Fix_Rate_example_combined}
\end{table*}

\begin{table*}[t]
  \caption{Pass rate of Compilation Errors for \framework{} using CodeLlama}
  \centering
  \renewcommand{\arraystretch}{1.2}
  \setlength{\tabcolsep}{5pt} 
{\small
  \begin{tabular}{l r@{/}r@{\hspace{4pt}}r r@{/}r@{\hspace{4pt}}r r@{/}r@{\hspace{4pt}}r r@{/}r@{\hspace{4pt}}r r@{/}r@{\hspace{4pt}}r}
  \toprule
  \textbf{Compilation error type} & \multicolumn{3}{c}{\textbf{OpenIPC}} & \multicolumn{3}{c}{\textbf{STM32}} & \multicolumn{3}{c}{\textbf{RTEMS}} & \multicolumn{3}{c}{\textbf{Zephyr}} & \multicolumn{3}{c}{\textbf{Total}} \tabularnewline
  \midrule
  Environment Setup Failure      & 4 & 4     & (100\,\%) & 16 & 23    & (70\,\%)  & 172 & 286  & (60\,\%)  & 75 & 100   & (75\,\%)  & 267 & 413   & (65\,\%) \tabularnewline
  Syntax Error                   & 2 & 2     & (100\,\%) & 44 & 80    & (55\,\%)  & 145 & 321  & (45\,\%)  & 155 & 258   & (60\,\%)  & 346 & 661   & (52\,\%) \tabularnewline
  Hardware Dependency Error      & 4 & 14    & (29\,\%)  & 119 & 341  & (35\,\%)  & 245 & 874  & (28\,\%)  & 453 & 1372  & (33\,\%)  & 821 & 2601  & (32\,\%) \tabularnewline
  Non-hardware Dependency Error  & 3 & 11    & (27\,\%)  & \multicolumn{3}{c}{--}      & 100 & 249  & (40\,\%)  & 86 & 179    & (48\,\%)  & 189 & 439   & (43\,\%) \tabularnewline
  Compiler Configuration Error   & \multicolumn{3}{c}{--}      & \multicolumn{3}{c}{--}      & 16 & 54     & (30\,\%)  & 30 & 80     & (38\,\%)  & 46 & 134    & (34\,\%) \tabularnewline
  \bottomrule
  \end{tabular}
}
  \label{tab:compilationPrevalenceFixRate}
\end{table*}

Table~\ref{tab:Fix_Rate_example_combined} presents the results under random selection of human fix examples, project-specific fix examples, and using fix examples from other projects. We observe variation in pass rates across projects and models. Overall, CodeLlama achieves the highest pass rates, e.g., 41\,\% for Zephyr, which is closely followed by CodeT5+. Falcon and Bloom perform lower, particularly on RTEMS and OpenIPC. These findings align with prior studies~\cite{DBLP:journals/corr/abs-2308-12950,DBLP:conf/emnlp/WangLGB0H23}, which show that CodeLlama and CodeT5+ consistently outperform other models in code understanding and repair tasks, demonstrating stronger alignment with source-level programming patterns. 

To assess whether the choice of human-fix examples in the prompt influences repair success, we applied a Friedman test on the pass rates of \textit{CodeLlama} across three conditions: examples from the same project, from any project, and from other projects—measured over four OSS projects. The test revealed a statistically significant difference among the conditions ($\chi^2(2) = 6.62,\ p = 0.0366$), indicating that the source of human-fix examples has a measurable effect on model performance. Project-specific examples yield higher repair success rates, suggesting that contextual alignment between example fixes and project conventions enhances LLMs’ ability to generate correct and compilable repairs.

Compared to random selection, CodeLlama’s pass rate increases by 4–6 percentage points, while CodeT5+ and the smaller models show more modest gains. This indicates that project-specific examples improve model accuracy by providing contextual consistency with each project’s coding style and structure. Zephyr achieves the highest pass rate (45\,\%) with CodeLlama, highlighting the benefit of in-project contextual information.

To assess whether the choice of LLMs influences repair success, we applied a Chi-square test on the three sets of Table~\ref{tab:Fix_Rate_example_combined} across three conditions—examples from the same project, from any project, and from other projects—measured over four OSS projects. We observe Chi-square statistics are ($\chi^2(2) = 140.1368,\ p = 0.0000$). The results show that the CI pass rate depends on the LLM. These findings indicate that the differences in pass rates among the evaluated LLMs are statistically significant across the examined CI projects.

Finally, Table~\ref{tab:compilationPrevalenceFixRate} presents CodeLlama’s pass rate by compilation error type. Although hardware dependency errors are the most prevalent type of compilation errors, they have the lowest pass rate of 31\,\% on average. This is because these errors often involve complex interactions between code and hardware configurations, making them harder to diagnose and fix automatically. In contrast, syntax errors have a higher pass rate of 65\,\%, as they typically involve straightforward corrections to code structure or formatting. It indicates a promising avenue for future research to enhance LLM capabilities in understanding and resolving hardware-related issues, potentially through specialized training or incorporating hardware context into the repair process.   

To assess whether the repair rate depends on the type of build failure encountered, a Chi-square test of independence was performed using the data in Table~\ref{tab:compilationPrevalenceFixRate}. The calculated test statistics ($\chi^2(2) = 226.1664,\ p = 0.0000$). Therefore, the results indicate that the repair rate is significantly dependent on the type of build failure. In particular, syntax-related and environment setup errors were substantially more likely to be repaired successfully than hardware- or configuration-related failures, suggesting that the model handles syntactic and environmental issues more effectively than domain-specific dependency problems.

Non-hardware dependency errors and compiler configuration errors also show moderate pass rates of 38\,\% and 34\,\%, respectively, indicating that, while challenging, they are still more amenable to automated repair than hardware-related issues. Environment setup failures have the highest pass rate of 76\,\%, likely because they often involve clear-cut issues such as missing dependencies or misconfigurations that can be easily identified and resolved.

Based on the results of the statistical tests, we observe that the choice of LLM has a statistically significant influence on the pass rate across OSS embedded projects, and the source of human-fix examples has a measurable effect. Moreover, the repair rate is found to be significantly dependent on the type of build failure encountered.

\resultbox{
    \textbf{Finding 3:} \framework{} successfully repairs up to 45\,\% of CI compilation failures across the targeted projects. The specific LLM and the source of human-fix examples have a measurable effect on the pass rate which is found to be significantly dependent on the type of build failure. 
}

\subsection{Fix Evaluation}\label{subsec:fix_evaluation}
\begin{figure*}[htbp]
    \centering  
    \includegraphics[scale=.55]{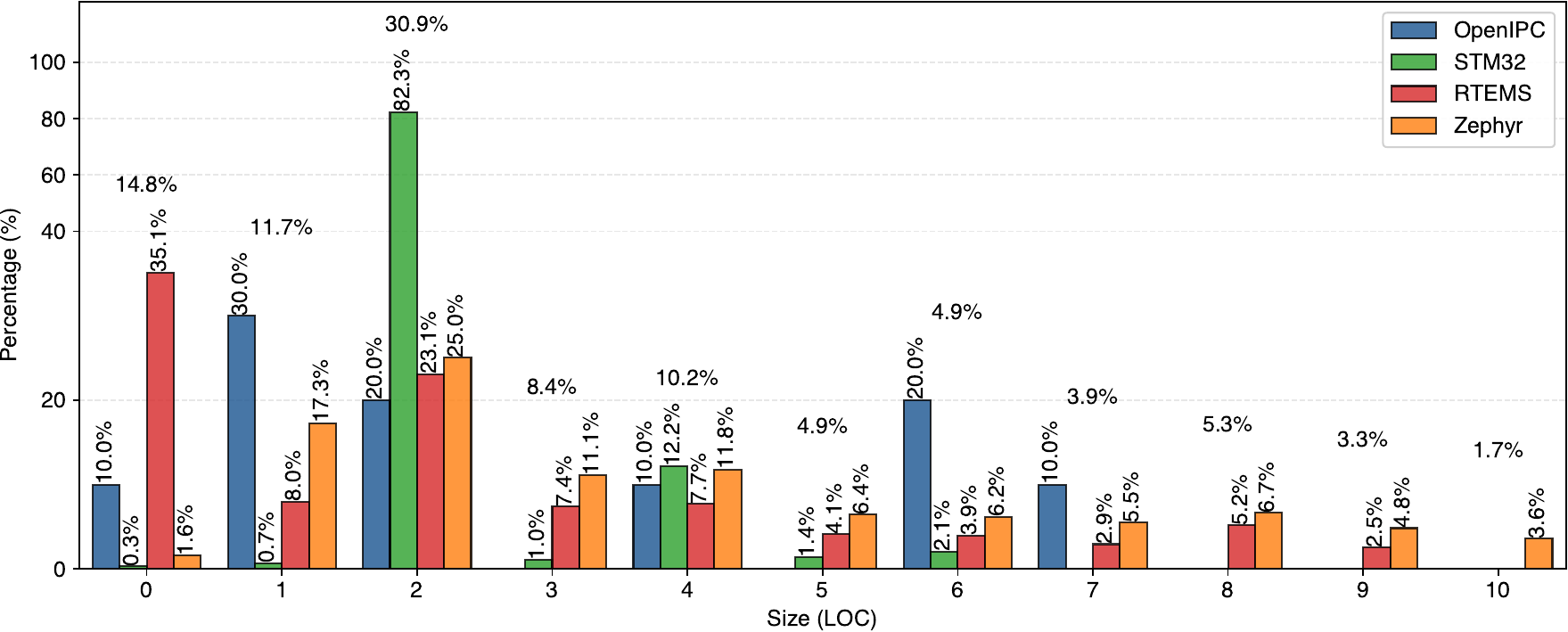}
    \Description{A diagram showing the resolution size of 4 different projects.}
    \caption{Resolution size for four projects}
    \label{fig:R_Size_4in1}
\end{figure*}

We further analyze the resolution size of each project based on CodeLlama, and analyze the fix code example. 
Fig.~\ref{fig:R_Size_4in1} illustrates the distribution of resolution sizes measured in lines of code (LOC). The most common resolution size is 2 LOC, accounting for 30.9\,\% of all fixes. STM32 has the most 2-LOC fixes (82.3\,\%). This suggests that most of the fixes involve one-line changes, which involve deleting one line and adding one line. 

The second most common size is 0 LOC, representing 14.8\,\% of fixes overall. While zero-line fixes may appear trivial, they often correspond to updates that affect binary artifacts or file permissions, which are not reflected in line-based metrics.

In such cases, a 0-LOC modification can still represent a valid and meaningful repair. However, RTEMS has the highest proportion of 0-LOC fixes (35.1\,\%), due to files being deleted or renamed, which is can detected and fixed by \framework{}. One example is that there is one guideline in the summary and log states that "MC68040 did not completely implement IEEE754 in hardware. MC68040 BSPs in RTEMS, this code can be removed.” This change results in the removal of all related files and a successful CI build later on, which is actually correct. This change involves 46 files, which in a considerable proportion of 0-LOC fixes. 

The majority of fixes lie within 0-2 LOC changes. This indicates that many compilation failures can be resolved with minimal code changes, often involving simple syntax corrections or minor adjustments to existing code. The remaining resolutions are distributed across larger sizes, with 2–5 LOC changes being the most frequent. \framework{} effectively addresses a range of fix sizes, demonstrating its capability to handle both small and moderately sized code modifications.

Many embedded software build failures stem from platform alias mismatches: code references a peripheral name that doesn’t exist on the selected board. The following are two examples of hardware dependency errors and non-hardware Dependency errors. 

\textbf{Fix example A} - Listing~\ref{ErrorLogA} shows a hardware dependency error where \texttt{LED\_BLUE} is undefined for the \texttt{NUCLEO-F411RE} board. The HAL for this platform provides \texttt{LED\_RED} but not \texttt{LED\_BLUE}, causing unresolved symbols. The fix in Listing~\ref{FixCodeA} replaces the invalid alias with a valid one, restoring successful compilation. This illustrates a common embedded compilation issue from mismatched code references and board-specific hardware definitions—frequent in open-source projects targeting multiple platforms—and highlights the need to distinguish such hardware abstraction mismatches from other build failures to improve automated detection and repair.

\begin{lstlisting}[style=mystyle, xleftmargin=1em, xrightmargin=0em, caption={Error log A}, label=ErrorLogA]
src/m.cpp:7:18: error: 'LED_BLUE' was not declared in this scope
DigitalOut led(LED_BLUE);
              ^
src/m.cpp:7:18: note: suggested alternative: 'LED_RED'
DigitalOut led(LED_BLUE);
              ^~~~~~~~
              LED_RED
*** [.pio/build/nucleo_f411re/src/m.cpp.o] Error 1
\end{lstlisting}

\begin{lstlisting}[style=mystyle, xleftmargin=1em, xrightmargin=0em, caption={Fix Code A}, label=FixCodeA]
- DigitalOut led(LED_BLUE);
+ DigitalOut led(LED_RED);
\end{lstlisting}

\textbf{Fix example B} - Listing~\ref{ErrorCodeB} shows a non-hardware dependency error caused by inconsistent function declarations: \texttt{MG\_ENABLE\_TCPIP} is declared to return \texttt{void} but defined to return \texttt{int}, producing a “conflicting types” error. The fix in Listing~\ref{FixCodeB} updates the prototype to match the definition, restoring consistency. Such errors often arise when implementations change without corresponding header updates, particularly in embedded and networked software where conditional compilation can obscure outdated prototypes. They represent software-level inconsistencies rather than hardware configuration issues and are common in large, evolving codebases.

\begin{lstlisting}[style=mystyle, xleftmargin=1em, xrightmargin=0em, caption={Erroneous code B}, label=ErrorCodeB]
// buggy code starts:
void mg_multicast_add(struct mg_connection *c, char *ip);
...
#if MG_ENABLE_TCPIP
int mg_multicast_add(struct mg_connection *c, char *ip) {
struct ip_mreq mreq;
// no include - type unknown
setsockopt(...);
// buggy code ends:
\end{lstlisting}

\begin{lstlisting}[style=mystyle, xleftmargin=0.6em, xrightmargin=0em, caption={Fix Code B}, label=FixCodeB]
- void mg_multicast_add(struct mg_connection *c, char *ip);
+ int mg_multicast_add(struct mg_connection *c, char *ip);
\end{lstlisting}

\resultbox{
    \textbf{Finding 4:} Most compilation errors from the targeted projects only require modifying up to 2 LOC, and \framework{} is capable of resolving a large portion of these errors.
}

\section{Threats to validity}\label{sec:threats2validity}
\noindent\textbf{Internal Validity} - Internal validity threats stem from differences in LLM training data, architecture, and error types that may influence repair outcomes. We mitigate this by evaluating four models, though each has inherent biases; specifically, LLMs may inadvertently introduce new bugs or security vulnerabilities during the repair process. Future work should explore a broader range of LLMs to reinforce reliability.

\noindent\textbf{External Validity} - External validity is supported by an analysis of four diverse open-source embedded projects, which offer a realistic view of CI repair behavior. While not fully generalizable to all domains, the use of off-the-shelf LLMs reflects practical deployment, and future studies can extend this by testing fine-tuned or domain-specific models across more varied settings.

\noindent\textbf{Construct Validity} - Construct validity is supported by using the CI pass rate as a practical and consistent indicator of repair success, reflecting build completion. However, this metric may limit \framework's ability to account for semantic correctness, potentially masking logical errors or vulnerabilities introduced by the LLM. Furthermore, restricting repairs to single files ensures reproducibility but limits the evaluation of \framework's effectiveness in resolving multiple simultaneous compilation errors that span across the project. Future work should incorporate semantic validation and multi-file repairs for greater realism.

\noindent\textbf{Conclusion Validity} - While the statistical tests confirm overall trends, they may miss finer interactions between model behavior and project characteristics. Expanding the dataset and applying more detailed analyses with confidence intervals would further strengthen the robustness of the findings.

\section{Conclusion}\label{sec:conclusion}
This study introduces \framework{}, a practical framework for integrating LLM-based automated program repair into embedded CI workflows across diverse platforms, including GitHub Actions and GitLab CI. It operates effectively within four open-source embedded projects—OpenIPC, STM32, RTEMS, and Zephyr—each employing distinct build systems. Within \framework{}, four open-source LLMs were evaluated based on accessibility, code-oriented pretraining, dataset diversity, and model size suitable for local inference. The framework enables automatic reproduction of failed builds and generation of targeted repair candidates with minimal human intervention.

Analyzing over~\num{10000} pull and merge requests, we observed build-stage failure rates of 7–11\,\%, reproducing \num{4248} compilation errors locally. Hardware and dependency issues dominated (71.5\,\%), compared to 37\,\% in industrial systems~\cite{fu2022prevalence}, indicating greater configuration diversity and less stable support in OSS ecosystems. Error distributions varied significantly across projects, suggesting project-specific CI failure patterns. Overall, most compilation failures stemmed from hardware or dependency issues rather than syntax errors.

Among the evaluated models, CodeLlama achieved the highest CI pass rate (45\,\%), followed by CodeT5+, particularly when prompts incorporated project-specific human-fix examples. Falcon and Bloom, lacking code-focused pretraining, performed less effectively, highlighting the benefit of domain specialization. Statistical analysis confirmed that both the choice of LLM and the source of fix examples significantly influenced repair success, and that syntax and environment errors were more reliably resolved than hardware or configuration failures. Most successful repairs modified only a few lines of code, demonstrating that many embedded build errors can be efficiently corrected through localized changes. Collectively, these results show that \framework{} provides an effective, scalable means of resolving CI build failures with minimal overhead.

Overall, \framework{} offers a favorable trade-off between setup effort and reward: once adapted to a project’s build system, it can automatically reproduce, analyze, and repair most CI compilation errors with minimal manual input. Adding a new project requires only modest configuration—primarily updating build scripts or dependency setup—while the resulting automation significantly reduces developer effort in diagnosing and fixing recurring CI build failures.

\section{Future work}\label{sec:futurework}
Future work will extend \framework{} in three key directions.  

First, we plan to expand its repair capability beyond compilation failures to include runtime failures that occur during CI builds. In addition to executing existing test suites, we will validate repaired programs using static analysis tools and runtime checkers (e.g., sanitizers or assertion-based instrumentation) to detect latent defects that may not be exposed by tests alone. Addressing these failure modes will further reduce manual intervention and improve overall build stability.

Second, we plan to extend \framework{} to address test failures that occur after successful compilation and execution, focusing on repairing the code under test rather than modifying the tests themselves. This would allow \framework{} to correct functional defects revealed by existing test suites, complementing its ability to handle build-time and runtime execution failures within the same CI run.

Finally, we intend to investigate domain adaptation of the underlying LLMs using project-specific build logs and historical repair data, including both lightweight parameter fine-tuning and prompt-based techniques. Such adaptation is expected to improve repair accuracy, reduce prompt sensitivity, and enable more consistent performance across heterogeneous embedded platforms.

\section*{Data Availability}
The datasets and statistical analysis scripts supporting the findings of this study are available on Zenodo at:  
\href{https://doi.org/10.5281/zenodo.18378245}{\underline{\textcolor{blue}{MSR-Repo}}}.

\begin{acks} 
This work was partially supported by the Wallenberg Artificial Intelligence, Autonomous Systems and Software Program (WASP) funded by the Knut and Alice Wallenberg Foundation.
\end{acks}

\bibliographystyle{ACM-Reference-Format}
\bibliography{references}

@InProceedings{	  fu2022prevalence,
  author	= {Han Fu and Sigrid Eldh and Kristian Wiklund and Andreas
		  Ermedahl and Cyrille Artho},
  title		= {Prevalence of continuous integration failures in
		  industrial systems with hardware-in-the-loop testing},
  booktitle	= {{IEEE} Int.\ Symposium on Software Reliability Engineering
		  Workshops, {ISSRE} - Workshops},
  pages		= {61--66},
  publisher	= {{IEEE}},
  address	= {Piscataway, NJ, USA},
  year		= {2022},
  url		= {https://doi.org/10.1109/ISSREW55968.2022.00040},
  doi		= {10.1109/ISSREW55968.2022.00040}
}

@InProceedings{	  fueweha24,
  author	= {Han Fu and Sigrid Eldh and Kristian Wiklund and Andreas
		  Ermedahl and Philipp Haller and Cyrille Artho},
  title		= {In industrial embedded software, are some compilation
		  errors easier to localize and fix than others?},
  booktitle	= {{IEEE} Conference on Software Testing, Verification and
		  Validation, {ICST}},
  pages		= {383--394},
  publisher	= {{IEEE}},
  year		= {2024},
  address	= {Piscataway, NJ, USA},
  url		= {https://doi.org/10.1109/ICST60714.2024.00042},
  doi		= {10.1109/ICST60714.2024.00042},
  bibsource	= {dblp computer science bibliography, https://dblp.org}
}

@InProceedings{	  dblp:conf/emnlp/wanglgb0h23,
  author	= {Yue Wang and Hung Le and Akhilesh Gotmare and Nghi D. Q.
		  Bui and Junnan Li and Steven C. H. Hoi},
  editor	= {Houda Bouamor and Juan Pino and Kalika Bali},
  title		= {CodeT5+: Open Code Large Language Models for Code
		  Understanding and Generation},
  booktitle	= {Proceedings of the 2023 Conference on Empirical Methods in
		  Natural Language Processing, {EMNLP} 2023, Singapore,
		  December 6-10, 2023},
  pages		= {1069--1088},
  publisher	= {Association for Computational Linguistics},
  year		= {2023},
  address	= {Singapore},
  url		= {https://doi.org/10.18653/v1/2023.emnlp-main.68},
  doi		= {10.18653/V1/2023.EMNLP-MAIN.68},
  bibsource	= {dblp computer science bibliography, https://dblp.org}
}

@Article{	  dblp:journals/corr/abs-2107-03374,
  author	= {Mark Chen and Jerry Tworek and Heewoo Jun and Qiming Yuan
		  and Henrique Pond{\'{e}} de Oliveira Pinto and Jared Kaplan
		  and Harri Edwards and Yuri Burda and Nicholas Joseph and
		  Greg Brockman and Alex Ray and Raul Puri and Gretchen
		  Krueger and Michael Petrov and Heidy Khlaaf and Girish
		  Sastry and Pamela Mishkin and Brooke Chan and Scott Gray
		  and Nick Ryder and Mikhail Pavlov and Alethea Power and
		  Lukasz Kaiser and Mohammad Bavarian and Clemens Winter and
		  Philippe Tillet and Felipe Petroski Such and Dave Cummings
		  and Matthias Plappert and Fotios Chantzis and Elizabeth
		  Barnes and Ariel Herbert{-}Voss and William Hebgen Guss and
		  Alex Nichol and Alex Paino and Nikolas Tezak and Jie Tang
		  and Igor Babuschkin and Suchir Balaji and Shantanu Jain and
		  William Saunders and Christopher Hesse and Andrew N. Carr
		  and Jan Leike and Joshua Achiam and Vedant Misra and Evan
		  Morikawa and Alec Radford and Matthew Knight and Miles
		  Brundage and Mira Murati and Katie Mayer and Peter Welinder
		  and Bob McGrew and Dario Amodei and Sam McCandlish and Ilya
		  Sutskever and Wojciech Zaremba},
  title		= {Evaluating Large Language Models Trained on Code},
  journal	= {CoRR},
  volume	= {abs/2107.03374},
  year		= {2021},
  url		= {https://arxiv.org/abs/2107.03374},
  eprinttype	= {arXiv},
  eprint	= {2107.03374},
  pages		= {},
  bibsource	= {dblp computer science bibliography, https://dblp.org}
}

@Article{	  dblp:journals/corr/abs-2305-02309,
  author	= {Erik Nijkamp and Hiroaki Hayashi and Caiming Xiong and
		  Silvio Savarese and Yingbo Zhou},
  title		= {CodeGen2: Lessons for Training LLMs on Programming and
		  Natural Languages},
  journal	= {CoRR},
  volume	= {abs/2305.02309},
  year		= {2023},
  url		= {https://doi.org/10.48550/arXiv.2305.02309},
  doi		= {10.48550/ARXIV.2305.02309},
  eprinttype	= {arXiv},
  eprint	= {2305.02309},
  pages		= {},
  bibsource	= {dblp computer science bibliography, https://dblp.org}
}

@Article{	  dblp:journals/corr/abs-2308-12950,
  author	= {Baptiste Rozi{\`{e}}re and Jonas Gehring and Fabian
		  Gloeckle and Sten Sootla and Itai Gat and Xiaoqing Ellen
		  Tan and Yossi Adi and Jingyu Liu and Tal Remez and
		  J{\'{e}}r{\'{e}}my Rapin and Artyom Kozhevnikov and Ivan
		  Evtimov and Joanna Bitton and Manish Bhatt and Cristian
		  Canton{-}Ferrer and Aaron Grattafiori and Wenhan Xiong and
		  Alexandre D{\'{e}}fossez and Jade Copet and Faisal Azhar
		  and Hugo Touvron and Louis Martin and Nicolas Usunier and
		  Thomas Scialom and Gabriel Synnaeve},
  title		= {Code Llama: Open Foundation Models for Code},
  journal	= {CoRR},
  volume	= {abs/2308.12950},
  year		= {2023},
  url		= {https://doi.org/10.48550/arXiv.2308.12950},
  doi		= {10.48550/ARXIV.2308.12950},
  eprinttype	= {arXiv},
  eprint	= {2308.12950},
  pages		= {},
  bibsource	= {dblp computer science bibliography, https://dblp.org}
}

@Article{	  dblp:journals/corr/abs-2401-14196,
  author	= {Daya Guo and Qihao Zhu and Dejian Yang and Zhenda Xie and
		  Kai Dong and Wentao Zhang and Guanting Chen and Xiao Bi and
		  Y. Wu and Y. K. Li and Fuli Luo and Yingfei Xiong and
		  Wenfeng Liang},
  title		= {DeepSeek-Coder: When the Large Language Model Meets
		  Programming - The Rise of Code Intelligence},
  journal	= {CoRR},
  volume	= {abs/2401.14196},
  year		= {2024},
  pages		= {},
  number	= {},
  url		= {https://doi.org/10.48550/arXiv.2401.14196},
  doi		= {10.48550/ARXIV.2401.14196},
  eprinttype	= {arXiv},
  eprint	= {2401.14196},
  bibsource	= {dblp computer science bibliography, https://dblp.org}
}

@Article{	  dblp:journals/tse/gouesnfw12,
  author	= {Claire {Le Goues} and ThanhVu Nguyen and Stephanie Forrest
		  and Westley Weimer},
  title		= {GenProg: {A} Generic Method for Automatic Software
		  Repair},
  journal	= {{IEEE} Trans. Software Eng.},
  volume	= {38},
  number	= {1},
  pages		= {54--72},
  year		= {2012},
  url		= {https://doi.org/10.1109/TSE.2011.104},
  doi		= {10.1109/TSE.2011.104},
  bibsource	= {dblp computer science bibliography, https://dblp.org}
}

@InProceedings{	  dblp:conf/icse/weimerngf09,
  author	= {Westley Weimer and ThanhVu Nguyen and Claire {Le Goues}
		  and Stephanie Forrest},
  title		= {Automatically finding patches using genetic programming},
  booktitle	= {31st International Conference on Software Engineering,
		  {ICSE} 2009, May 16-24, 2009, Vancouver, Canada,
		  Proceedings},
  pages		= {364--374},
  publisher	= {{IEEE}},
  year		= {2009},
  address	= {Piscataway, NJ, USA},
  url		= {https://doi.org/10.1109/ICSE.2009.5070536},
  doi		= {10.1109/ICSE.2009.5070536},
  bibsource	= {dblp computer science bibliography, https://dblp.org}
}

@Article{	  dblp:journals/tse/afzalmsbg21,
  author	= {Afsoon Afzal and Manish Motwani and Kathryn T. Stolee and
		  Yuriy Brun and Claire {Le Goues}},
  title		= {SOSRepair: Expressive Semantic Search for Real-World
		  Program Repair},
  journal	= {{IEEE} Trans. Software Eng.},
  volume	= {47},
  number	= {10},
  pages		= {2162--2181},
  year		= {2021},
  url		= {https://doi.org/10.1109/TSE.2019.2944914},
  doi		= {10.1109/TSE.2019.2944914},
  bibsource	= {dblp computer science bibliography, https://dblp.org}
}

@InProceedings{	  dblp:conf/icse/mechtaevyr16,
  author	= {Sergey Mechtaev and Jooyong Yi and Abhik Roychoudhury},
  editor	= {Laura K. Dillon and Willem Visser and Laurie A. Williams},
  title		= {Angelix: scalable multiline program patch synthesis via
		  symbolic analysis},
  booktitle	= {Proceedings of the 38th International Conference on
		  Software Engineering, {ICSE} 2016, Austin, TX, USA, May
		  14-22, 2016},
  pages		= {691--701},
  publisher	= {{ACM}},
  year		= {2016},
  address	= {New York, NY, USA},
  url		= {https://doi.org/10.1145/2884781.2884807},
  doi		= {10.1145/2884781.2884807},
  bibsource	= {dblp computer science bibliography, https://dblp.org}
}

@Article{	  dblp:journals/tse/chenkm23,
  author	= {Zimin Chen and Steve Kommrusch and Martin Monperrus},
  title		= {Neural Transfer Learning for Repairing Security
		  Vulnerabilities in {C} Code},
  journal	= {{IEEE} Trans. Software Eng.},
  volume	= {49},
  number	= {1},
  pages		= {147--165},
  year		= {2023},
  url		= {https://doi.org/10.1109/TSE.2022.3147265},
  doi		= {10.1109/TSE.2022.3147265},
  bibsource	= {dblp computer science bibliography, https://dblp.org}
}

@InProceedings{	  dblp:conf/icse/fangmrt23,
  author	= {Zhiyu Fan and Xiang Gao and Martin Mirchev and Abhik
		  Roychoudhury and Shin Hwei Tan},
  title		= {Automated Repair of Programs from Large Language Models},
  booktitle	= {45th {IEEE/ACM} International Conference on Software
		  Engineering, {ICSE} 2023, Melbourne, Australia, May 14-20,
		  2023},
  pages		= {1469--1481},
  publisher	= {{IEEE}},
  year		= {2023},
  address	= {Piscataway, NJ, USA},
  url		= {https://doi.org/10.1109/ICSE48619.2023.00128},
  doi		= {10.1109/ICSE48619.2023.00128},
  bibsource	= {dblp computer science bibliography, https://dblp.org}
}

@Article{	  dblp:journals/pacmse/hossain0z0clnt24,
  author	= {Soneya Binta Hossain and Nan Jiang and Qiang Zhou and
		  Xiaopeng Li and Wen{-}Hao Chiang and Yingjun Lyu and Hoan
		  Anh Nguyen and Omer Tripp},
  title		= {A Deep Dive into Large Language Models for Automated Bug
		  Localization and Repair},
  journal	= {Proc. {ACM} Softw. Eng.},
  volume	= {1},
  number	= {{FSE}},
  pages		= {1471--1493},
  year		= {2024},
  url		= {https://doi.org/10.1145/3660773},
  doi		= {10.1145/3660773},
  bibsource	= {dblp computer science bibliography, https://dblp.org}
}

@Misc{		  sahoo2025systematicsurveypromptengineering,
  title		= {A Systematic Survey of Prompt Engineering in Large
		  Language Models: Techniques and Applications},
  author	= {Pranab Sahoo and Ayush Kumar Singh and Sriparna Saha and
		  Vinija Jain and Samrat Mondal and Aman Chadha},
  year		= {2025},
  eprint	= {2402.07927},
  archiveprefix	= {arXiv},
  primaryclass	= {cs.AI},
  url		= {https://arxiv.org/abs/2402.07927}
}

@InProceedings{	  dblp:conf/msr/rauschhl017,
  author	= {Thomas Rausch and Waldemar Hummer and Philipp Leitner and
		  Stefan Schulte},
  editor	= {Jes{\'{u}}s M. Gonz{\'{a}}lez{-}Barahona and Abram Hindle
		  and Lin Tan},
  title		= {An empirical analysis of build failures in the continuous
		  integration workflows of Java-based open-source software},
  booktitle	= {Proceedings of the 14th International Conference on Mining
		  Software Repositories, {MSR} 2017, Buenos Aires, Argentina,
		  May 20-28, 2017},
  pages		= {345--355},
  publisher	= {{IEEE} Computer Society},
  year		= {2017},
  address	= {Piscataway, NJ, USA},
  url		= {https://doi.org/10.1109/MSR.2017.54},
  doi		= {10.1109/MSR.2017.54},
  bibsource	= {dblp computer science bibliography, https://dblp.org}
}

@Article{	  mchugh2013chi,
  title		= {The chi-square test of independence},
  author	= {McHugh, Mary L},
  journal	= {Biochemia medica},
  volume	= {23},
  number	= {2},
  pages		= {143--149},
  year		= {2013},
  publisher	= {Medicinska naklada}
}

@Article{	  friedman1937use,
  title		= {The use of ranks to avoid the assumption of normality
		  implicit in the analysis of variance},
  author	= {Friedman, Milton},
  journal	= {Journal of the american statistical association},
  volume	= {32},
  number	= {200},
  pages		= {675--701},
  year		= {1937},
  publisher	= {Taylor \& Francis}
}

@Article{	  dblp:journals/corr/abs-2303-18223,
  author	= {Wayne Xin Zhao and Kun Zhou and Junyi Li and Tianyi Tang
		  and Xiaolei Wang and Yupeng Hou and Yingqian Min and
		  Beichen Zhang and Junjie Zhang and Zican Dong and Yifan Du
		  and Chen Yang and Yushuo Chen and Zhipeng Chen and Jinhao
		  Jiang and Ruiyang Ren and Yifan Li and Xinyu Tang and
		  Zikang Liu and Peiyu Liu and Jian{-}Yun Nie and Ji{-}Rong
		  Wen},
  title		= {A survey of large language models},
  journal	= {CoRR},
  volume	= {abs/2303.18223},
  year		= {2023},
  pages		= {1--140},
  numpages	= {140},
  url		= {https://doi.org/10.48550/arXiv.2303.18223},
  doi		= {10.48550/ARXIV.2303.18223},
  eprinttype	= {arXiv},
  eprint	= {2303.18223},
  bibsource	= {dblp computer science bibliography, https://dblp.org}
}

@Article{	  wei2022chain,
  title		= {Chain-of-thought prompting elicits reasoning in large
		  language models},
  author	= {Wei, Jason and Wang, Xuezhi and Schuurmans, Dale and
		  Bosma, Maarten and Xia, Fei and Chi, Ed and Le, Quoc V and
		  Zhou, Denny and others},
  journal	= {Advances in Neural Information Processing Systems},
  volume	= {35},
  pages		= {24824--24837},
  year		= {2022}
}

@Article{	  panlwcww24,
  author	= {Shirui Pan and Linhao Luo and Yufei Wang and Chen Chen and
		  Jiapu Wang and Xindong Wu},
  title		= {Unifying large language models and knowledge graphs: {A}
		  roadmap},
  journal	= {{IEEE} Trans. Knowl. Data Eng.},
  volume	= {36},
  number	= {7},
  pages		= {3580--3599},
  year		= {2024},
  url		= {https://doi.org/10.1109/TKDE.2024.3352100},
  doi		= {10.1109/TKDE.2024.3352100},
  bibsource	= {dblp computer science bibliography, https://dblp.org}
}

@InProceedings{	  kerzazi2014automated,
  author	= {Noureddine Kerzazi and Foutse Khomh and Bram Adams},
  title		= {Why Do Automated Builds Break? An Empirical Study},
  booktitle	= {30th {IEEE} International Conference on Software
		  Maintenance and Evolution, Victoria, BC, Canada, September
		  29 - October 3, 2014},
  pages		= {41--50},
  publisher	= {{IEEE} Computer Society},
  year		= {2014},
  address	= {Piscataway, NJ, USA},
  url		= {https://doi.org/10.1109/ICSME.2014.26},
  doi		= {10.1109/ICSME.2014.26},
  bibsource	= {dblp computer science bibliography, https://dblp.org}
}

@InProceedings{	  fischer2020forgotten,
  author	= {Anders Fischer{-}Nielsen and Zhoulai Fu and Ting Su and
		  Andrzej Wasowski},
  editor	= {Gregg Rothermel and Doo{-}Hwan Bae},
  title		= {The forgotten case of the dependency bugs: on the example
		  of the robot operating system},
  booktitle	= {{ICSE-SEIP} 2020: 42nd International Conference on
		  Software Engineering, Software Engineering in Practice,
		  Seoul, South Korea, 27 June - 19 July, 2020},
  pages		= {21--30},
  publisher	= {{ACM}},
  year		= {2020},
  address	= {New York, NY, USA},
  url		= {https://doi.org/10.1145/3377813.3381364},
  doi		= {10.1145/3377813.3381364},
  bibsource	= {dblp computer science bibliography, https://dblp.org}
}

@InProceedings{	  khazem2018making,
  author	= {Kareem Khazem and Earl T. Barr and Petr Hosek},
  editor	= {Frank Tip and Eric Bodden},
  title		= {Making data-driven porting decisions with {Tuscan}},
  booktitle	= {Proceedings of the 27th {ACM} {SIGSOFT} International
		  Symposium on Software Testing and Analysis, {ISSTA} 2018,
		  Amsterdam, The Netherlands, July 16-21, 2018},
  pages		= {276--286},
  publisher	= {{ACM}},
  year		= {2018},
  address	= {New York, NY, USA},
  url		= {https://doi.org/10.1145/3213846.3213855},
  doi		= {10.1145/3213846.3213855},
  bibsource	= {dblp computer science bibliography, https://dblp.org}
}

@InProceedings{	  zakaria2022mapping,
  author	= {Farid Zakaria and Thomas R. W. Scogland and Todd Gamblin
		  and Carlos Maltzahn},
  editor	= {Felix Wolf and Sameer Shende and Candace Culhane and Sadaf
		  R. Alam and Heike Jagode},
  title		= {Mapping Out the {HPC} Dependency Chaos},
  booktitle	= {{SC22:} International Conference for High Performance
		  Computing, Networking, Storage and Analysis, Dallas, TX,
		  USA, November 13-18, 2022},
  pages		= {34:1--34:12},
  publisher	= {{IEEE}},
  year		= {2022},
  address	= {Piscataway, NJ, USA},
  url		= {https://doi.org/10.1109/SC41404.2022.00039},
  doi		= {10.1109/SC41404.2022.00039},
  bibsource	= {dblp computer science bibliography, https://dblp.org}
}

@InProceedings{	  ratti2018conceptual,
  author	= "Ratti, Nisha and Kaur, Parminder",
  editor	= "Bhatia, Sanjiv K. and Mishra, Krishn K. and Tiwari,
		  Shailesh and Singh, Vivek Kumar",
  title		= "A Conceptual Framework for Analysing the Source Code
		  Dependencies",
  booktitle	= "Advances in Computer and Computational Sciences",
  year		= "2018",
  publisher	= "Springer Singapore",
  address	= "Singapore",
  pages		= "333--341",
  isbn		= "978-981-10-3773-3"
}

@InProceedings{	  rodrigues2021clp,
  author	= {Kirk Rodrigues and Yu Luo and Ding Yuan},
  editor	= {Angela Demke Brown and Jay R. Lorch},
  title		= {{CLP:} Efficient and Scalable Search on Compressed Text
		  Logs},
  booktitle	= {15th {USENIX} Symposium on Operating Systems Design and
		  Implementation, {OSDI} 2021, July 14-16, 2021},
  pages		= {183--198},
  publisher	= {{USENIX} Association},
  year		= {2021},
  address	= {Berkeley, CA, USA},
  url		= {https://www.usenix.org/conference/osdi21/presentation/rodrigues},
  bibsource	= {dblp computer science bibliography, https://dblp.org}
}

@Article{	  dai2020logram,
  author	= {Hetong Dai and Heng Li and Che{-}Shao Chen and Weiyi Shang
		  and Tse{-}Hsun Chen},
  title		= {Logram: Efficient Log Parsing Using $n$n-Gram
		  Dictionaries},
  journal	= {{IEEE} Trans. Software Eng.},
  volume	= {48},
  number	= {3},
  pages		= {879--892},
  year		= {2022},
  url		= {https://doi.org/10.1109/TSE.2020.3007554},
  doi		= {10.1109/TSE.2020.3007554},
  bibsource	= {dblp computer science bibliography, https://dblp.org}
}

@InProceedings{	  du2017deeplog,
  author	= {Min Du and Feifei Li and Guineng Zheng and Vivek
		  Srikumar},
  editor	= {Bhavani Thuraisingham and David Evans and Tal Malkin and
		  Dongyan Xu},
  title		= {DeepLog: {Anomaly} Detection and Diagnosis from System
		  Logs through Deep Learning},
  booktitle	= {Proceedings of the 2017 {ACM} {SIGSAC} Conference on
		  Computer and Communications Security, {CCS} 2017, Dallas,
		  TX, USA, October 30 - November 03, 2017},
  pages		= {1285--1298},
  address	= {New York, NY, USA},
  publisher	= {{ACM}},
  year		= {2017},
  url		= {https://doi.org/10.1145/3133956.3134015},
  doi		= {10.1145/3133956.3134015},
  bibsource	= {dblp computer science bibliography, https://dblp.org}
}

@InProceedings{	  zhu2019tools,
  author	= {Jieming Zhu and Shilin He and Jinyang Liu and Pinjia He
		  and Qi Xie and Zibin Zheng and Michael R. Lyu},
  editor	= {Helen Sharp and Mike Whalen},
  title		= {Tools and benchmarks for automated log parsing},
  booktitle	= {Proceedings of the 41st International Conference on
		  Software Engineering: Software Engineering in Practice,
		  {ICSE} {(SEIP)} 2019, Montreal, QC, Canada, May 25-31,
		  2019},
  pages		= {121--130},
  publisher	= {{IEEE} / {ACM}},
  year		= {2019},
  address	= {Piscataway, NJ, USA},
  url		= {https://doi.org/10.1109/ICSE-SEIP.2019.00021},
  doi		= {10.1109/ICSE-SEIP.2019.00021},
  bibsource	= {dblp computer science bibliography, https://dblp.org}
}

@InProceedings{	  fu2014digging,
  author	= {Xiaoyu Fu and Rui Ren and Sally A. McKee and Jianfeng Zhan
		  and Ninghui Sun},
  title		= {Digging deeper into cluster system logs for failure
		  prediction and root cause diagnosis},
  booktitle	= {2014 {IEEE} International Conference on Cluster Computing,
		  {CLUSTER} 2014, Madrid, Spain, September 22-26, 2014},
  pages		= {103--112},
  publisher	= {{IEEE} Computer Society},
  year		= {2014},
  address	= {Piscataway, NJ, USA},
  url		= {https://doi.org/10.1109/CLUSTER.2014.6968768},
  doi		= {10.1109/CLUSTER.2014.6968768},
  bibsource	= {dblp computer science bibliography, https://dblp.org}
}

@InProceedings{	  he2018identifying,
  author	= {Shilin He and Qingwei Lin and Jian{-}Guang Lou and Hongyu
		  Zhang and Michael R. Lyu and Dongmei Zhang},
  editor	= {Gary T. Leavens and Alessandro Garcia and Corina S.
		  Pasareanu},
  title		= {Identifying impactful service system problems via log
		  analysis},
  booktitle	= {Proceedings of the 2018 {ACM} Joint Meeting on European
		  Software Engineering Conference and Symposium on the
		  Foundations of Software Engineering, {ESEC/SIGSOFT} {FSE}
		  2018, Lake Buena Vista, FL, USA, November 04-09, 2018},
  pages		= {60--70},
  publisher	= {{ACM}},
  year		= {2018},
  address	= {New York, NY, USA},
  url		= {https://doi.org/10.1145/3236024.3236083},
  doi		= {10.1145/3236024.3236083},
  bibsource	= {dblp computer science bibliography, https://dblp.org}
}

@InProceedings{	  zhang2021onion,
  author	= {Xu Zhang and Yong Xu and Si Qin and Shilin He and Bo Qiao
		  and Ze Li and Hongyu Zhang and Xukun Li and Yingnong Dang
		  and Qingwei Lin and Murali Chintalapati and Saravanakumar
		  Rajmohan and Dongmei Zhang},
  editor	= {Diomidis Spinellis and Georgios Gousios and Marsha Chechik
		  and Massimiliano Di Penta},
  title		= {Onion: identifying incident-indicating logs for cloud
		  systems},
  booktitle	= {{ESEC/FSE} 2021: 29th {ACM} Joint European Software
		  Engineering Conference and Symposium on the Foundations of
		  Software Engineering, Athens, Greece, August 23-28, 2021},
  pages		= {1253--1263},
  publisher	= {{ACM}},
  address	= {New York, NY, USA},
  year		= {2021},
  url		= {https://doi.org/10.1145/3468264.3473919},
  doi		= {10.1145/3468264.3473919},
  bibsource	= {dblp computer science bibliography, https://dblp.org}
}

@InProceedings{	  olsson2012climbing,
  title		= {Climbing the "Stairway to Heaven"---A Multiple-Case Study
		  Exploring Barriers in the Transition from Agile Development
		  towards Continuous Deployment of Software},
  author	= {Olsson, Helena Holmstr{\"o}m and Alahyari, Hiva and Bosch,
		  Jan},
  booktitle	= {38th Euromicro Conference on Software Engineering and
		  Advanced Applications},
  pages		= {392--399},
  year		= {2012},
  address	= {Piscataway, NJ, USA},
  publisher	= {IEEE},
  organization	= {IEEE}
}

@InProceedings{	  lwakatare2016towards,
  title		= {Towards {DevOps} in the embedded systems domain: Why is it
		  so hard?},
  author	= {Lwakatare, Lucy Ellen and Karvonen, Teemu and Sauvola,
		  Tanja and Kuvaja, Pasi and Olsson, Helena Holmstr{\"o}m and
		  Bosch, Jan and Oivo, Markku},
  booktitle	= {49th Hawaii International Conference on System Sciences
		  (HICSS)},
  pages		= {5437--5446},
  year		= {2016},
  publisher	= {IEEE},
  address	= {Piscataway, NJ, USA},
  organization	= {IEEE}
}

@InProceedings{	  dblp:conf/wcre/golzadehdm22,
  author	= {Mehdi Golzadeh and Alexandre Decan and Tom Mens},
  title		= {On the rise and fall of {CI} services in GitHub},
  booktitle	= {{IEEE} International Conference on Software Analysis,
		  Evolution and Reengineering, {SANER} 2022, Honolulu, HI,
		  USA, March 15-18, 2022},
  pages		= {662--672},
  publisher	= {{IEEE}},
  year		= {2022},
  address	= {Piscataway, NJ, USA},
  url		= {https://doi.org/10.1109/SANER53432.2022.00084},
  doi		= {10.1109/SANER53432.2022.00084},
  bibsource	= {dblp computer science bibliography, https://dblp.org}
}

@InProceedings{	  dblp:conf/icse/zhugfr23,
  author	= {Hao{-}Nan Zhu and Kevin Z. Guan and Robert M. Furth and
		  Cindy Rubio{-}Gonz{\'{a}}lez},
  title		= {Actionsremaker: Reproducing {GITHUB} Actions},
  booktitle	= {45th {IEEE/ACM} International Conference on Software
		  Engineering: {ICSE} 2023 Companion Proceedings, Melbourne,
		  Australia, May 14-20, 2023},
  pages		= {11--15},
  publisher	= {{IEEE}},
  address	= {Piscataway, NJ, USA},
  year		= {2023},
  url		= {https://doi.org/10.1109/ICSE-Companion58688.2023.00015},
  doi		= {10.1109/ICSE-COMPANION58688.2023.00015},
  bibsource	= {dblp computer science bibliography, https://dblp.org}
}

@InProceedings{	  dblp:conf/msr/valenzuelatoledobkn23,
  author	= {Pablo Valenzuela{-}Toledo and Alexandre Bergel and Timo
		  Kehrer and Oscar Nierstrasz},
  title		= {{EGAD:} {A} moldable tool for GitHub Action analysis},
  booktitle	= {20th {IEEE/ACM} International Conference on Mining
		  Software Repositories, {MSR} 2023, Melbourne, Australia,
		  May 15-16, 2023},
  pages		= {260--264},
  publisher	= {{IEEE}},
  address	= {Piscataway, NJ, USA},
  year		= {2023},
  url		= {https://doi.org/10.1109/MSR59073.2023.00044},
  doi		= {10.1109/MSR59073.2023.00044},
  bibsource	= {dblp computer science bibliography, https://dblp.org}
}

@InProceedings{	  dblp:conf/icse/saavedrasm24,
  author	= {Nuno Saavedra and Andr{\'{e}} Silva and Martin Monperrus},
  title		= {GitBug-Actions: Building Reproducible Bug-Fix Benchmarks
		  with GitHub Actions},
  booktitle	= {Proceedings of the 2024 {IEEE/ACM} 46th International
		  Conference on Software Engineering: Companion Proceedings,
		  {ICSE} Companion 2024, Lisbon, Portugal, April 14-20,
		  2024},
  pages		= {1--5},
  publisher	= {{ACM}},
  year		= {2024},
  address	= {New York, NY, USA},
  url		= {https://doi.org/10.1145/3639478.3640023},
  doi		= {10.1145/3639478.3640023},
  bibsource	= {dblp computer science bibliography, https://dblp.org}
}

@Misc{		  gitlab_ci_docs_2024,
  author	= {{GitLab Documentation}},
  title		= {{GitLab CI/CD Pipelines}},
  year		= {2024},
  organization	= {GitLab B.V.},
  url		= {https://docs.gitlab.com/ee/ci/pipelines/},
  note		= {Accessed: 2024-11-10}
}

@Misc{		  github_actions_about_2024,
  author	= {{GitHub Documentation}},
  title		= {About {GitHub Actions}},
  year		= {2024},
  organization	= {GitHub, Inc.},
  url		= {https://docs.github.com/en/actions/learn-github-actions/understanding-github-actions},
  note		= {Accessed: 2024-11-10}
}

@Misc{		  fu2025autorepairtestcasesllms,
  title		= {Auto-repair without test cases: How LLMs fix compilation
		  errors in large industrial embedded code},
  author	= {Han Fu and Sigrid Eldh and Kristian Wiklund and Andreas
		  Ermedahl and Philipp Haller and Cyrille Artho},
  year		= {2025},
  eprint	= {2510.13575},
  archiveprefix	= {arXiv},
  primaryclass	= {cs.SE},
  url		= {https://arxiv.org/abs/2510.13575}
}

@InProceedings{	  dblp:conf/sigsoft/0001rjga19,
  author	= {Ali Mesbah and Andrew Rice and Emily Johnston and Nick
		  Glorioso and Edward Aftandilian},
  editor	= {Marlon Dumas and Dietmar Pfahl and Sven Apel and
		  Alessandra Russo},
  title		= {DeepDelta: learning to repair compilation errors},
  booktitle	= {Proceedings of the {ACM} Joint Meeting on European
		  Software Engineering Conference and Symposium on the
		  Foundations of Software Engineering, {ESEC/SIGSOFT} {FSE}
		  2019, Tallinn, Estonia, August 26-30, 2019},
  pages		= {925--936},
  publisher	= {{ACM}},
  address	= {New York, NY, USA},
  year		= {2019},
  url		= {https://doi.org/10.1145/3338906.3340455},
  doi		= {10.1145/3338906.3340455},
  bibsource	= {dblp computer science bibliography, https://dblp.org}
}

\end{document}